\documentclass[%
reprint,
superscriptaddress,
preprintnumbers,
nofootinbib,
 amsmath,amssymb,
 aps,
]{revtex4-1}

\usepackage[normalem]{ulem}
\usepackage[colorlinks]{hyperref}
\usepackage{graphicx}
\usepackage{dcolumn}
\usepackage{bm}
\usepackage{color}
\usepackage{bbold}
\usepackage[dvipsnames]{xcolor}
\usepackage{placeins}
\usepackage{slashed}
\usepackage{multirow}
\usepackage{subfigure}
\usepackage{tabularx}
\usepackage{mathtools}
\usepackage{booktabs}
\usepackage{blindtext}
\usepackage{soul}
\usepackage{slashed}
\usepackage{amsmath}
\usepackage{tikz}
\usepackage{comment}

\usetikzlibrary{decorations.pathmorphing}
\usetikzlibrary{arrows}
\usepackage{circuitikz}
\usetikzlibrary{arrows.meta}
\usetikzlibrary{shapes.misc}
\usetikzlibrary{positioning}
\usetikzlibrary{decorations.markings}

\usepackage[compat=1.1.0]{tikz-feynman}

\hypersetup{
  colorlinks=true,
  linkcolor=red,
  citecolor=blue,
  urlcolor=blue
}

\definecolor{coral}{RGB}{255,127,80}
\definecolor{indigo}{RGB}{75,0,130}
\definecolor{red}{rgb}{0.9, 0,0}
\definecolor{cerulean}{rgb}{0., 0.62,0.9}
\definecolor{navy}{rgb}{0.05, 0.05,0.8}

\newcommand{\GF}{G_\text{F}}

\newcommand{\MeV}{{\rm \, MeV}}

\def\refjnl#1{{\rm{#1}}}
\def\apj{\refjnl{ApJ}}                 
\def\apjl{\refjnl{ApJ}}                
\def\prc{\refjnl{Phys.~Rev.~C}}        
\def\prd{\refjnl{Phys.~Rev.~D}}        
\def\prl{\refjnl{Phys.~Rev.~Lett.}}    

\begin{document}
\begin{flushleft}
LAPTH-035/26
\end{flushleft}

\title{\texorpdfstring{$\Lambda$}{Lambda} hyperons in core-collapse supernovae: Equilibration and neutrino opacities}
\author{Ruben Zatini}
\email{ruben.zatini@iac.es}
\affiliation{Instituto de Astrof\'isica de Canarias,  C/ V\'ia L\'actea, s/n E38205 - La Laguna, Tenerife, Spain}
\affiliation{Universidad de La Laguna, Departamento de Astrof\'isica, La Laguna, Tenerife, Spain}
\affiliation{LAPTh, CNRS,  USMB, F-74940 Annecy, France}
\author{Jorge Martin Camalich}
\email{jcamalich@iac.es}
\affiliation{Instituto de Astrof\'isica de Canarias,  C/ V\'ia L\'actea, s/n E38205 - La Laguna, Tenerife, Spain}
\affiliation{Universidad de La Laguna, Departamento de Astrof\'isica, La Laguna, Tenerife, Spain}
\affiliation{CERN, Theoretical Physics Department, CH-1211 Geneva 23, Switzerland}
\author{Pasquale Dario Serpico}\email{serpico@lapth.cnrs.fr}
\affiliation{LAPTh, CNRS,  USMB, F-74940 Annecy, France}
\author{Tobias~Fischer}
\email{bert-tobias.fischer@pwr.edu.pl}
\affiliation{Institute of Theoretical Physics, Wrocław University of Science and Technology, Wybrzeże Wyspiańskiego 27, 50-370 Wrocław, Poland}
\affiliation{Research Center for Computational Physics and Data Processing, Institute of Physics, Silesian University in Opava, Bezručovo nám. 13, CZ-746-01 Opava, Czech Republic}
\date{\today}

\begin{abstract}
Strange hadrons are commonly included in dense-matter equation-of-state models by imposing chemical equilibrium, but the weak-interaction timescales required to establish it in core-collapse supernovae have not been systematically assessed. 
In this paper we compute the $\Lambda$-hyperon production rates in the hot, dense, and isospin-asymmetric conditions characteristic of post-collapse proto-neutron stars.
We find that local $\Lambda$ chemical equilibration is driven by nonleptonic strangeness-changing reactions, especially $NN\leftrightarrow N\Lambda$ scattering, on timescales of order $10^{-11}$--$10^{-10}$~s, many orders of magnitude shorter than macroscopic proto-neutron-star evolution timescales.
Using an effective-field-theory framework constrained by hypernuclear weak-decay data, we find that short-range contact interactions dominate the nonleptonic rates, beyond a pure one-meson-exchange description. Semileptonic channels are too slow to set the equilibrium $\Lambda$ abundance, but they open additional absorption channels for low-energy muon neutrinos and antineutrinos,  such as $\nu_\mu+\Lambda\to\mu^-+p$ and $p+\mu^-+\bar\nu_\mu\to\Lambda$. At low energies, these $\Lambda$-induced neutrino opacities exceed the corresponding nucleonic contributions for muon (anti)neutrinos, possibly influencing the evolution of the muon lepton number during proto-neutron-star deleptonization. These results support local chemical equilibrium for $\Lambda$ hyperons under the conditions studied and provide new weak-interaction input for flavor-dependent neutrino transport, muonization, and proto-neutron-star evolution.

\end{abstract}
\maketitle

\section{Introduction}
\label{sec:intro}

Massive stars with masses roughly above $9\,M_\odot$ end their lives in a core-collapse supernova (CCSN), triggered when the stellar core loses pressure support and collapses to densities exceeding nuclear saturation. At this point, the collapse of the inner core halts and reverses, generating a shock wave that propagates outward through the still-infalling stellar material. The shock eventually stalls due to energy losses from the dissociation of iron-group nuclei and from the prompt neutrino burst produced by electron captures on the newly liberated protons when the shock reaches the neutrinosphere.

The stellar core collapse converts roughly $\sim 10^{53}$~erg of gravitational energy into thermal energy of the proto-neutron star (PNS) that is  subsequently emitted as neutrinos. Neutrino emission from the PNS and the resulting heating in the post-shock gain region remain the leading explanation for shock revival~\cite{Bethe:1985sox}.  Alternative scenarios include the magneto-rotational mechanism~\cite{Bisnovatyi-Kogan70,LeBlancWilson70} and explosions triggered via a sufficiently strong first-order QCD phase transition~\cite{TakaharaSato1988PThPh80,Sagert09,Fischer18}.
The neutrino-heating mechanism, however, requires multi-dimensional simulations which show enhanced neutrino-heating efficiency due to convection in the post-shock layer and at high densities inside the PNS, stellar rotation as well as the presence of hydrodynamic instabilities ~(for recent reviews, see Refs.~ \cite{Janka:2006fh,Mirizzi:2015eza,Muller:2016izw,Janka2025ARNPS}).

The CCSN environment is characterized by extreme conditions: baryon densities exceeding few times nuclear saturation density, temperatures reaching up to 100~MeV~\cite{Burrows2026arXiv260209025R}, and large isospin asymmetry characterized by the hadronic charge fraction, $Y_Q$, which equals the proton abundance, $Y_Q=Y_p$, in the absence of other charged hadrons.
These conditions not only probe QCD matter at high baryon density~\cite{Fischer:2017zcr,Oertel:2016bki} but also allow for the appearance of heavy-flavor particles such as muons and hyperons. Muons can be produced once the electron chemical potential and the temperature become sufficiently large, providing the energy available in particle collisions to overcome the muon rest mass. Hyperons, in turn, are heavier strange baryons also expected to appear at high baryon densities and temperatures in the core of the PNS~\cite{Fortin:2017dsj,Camalich:2020wac,Kochankovski:2023trc,Fischer:2024ivh}.

Detailed CCSN simulations, implementing six-species Boltzmann neutrino transport and a comprehensive set of muonic weak processes, have shown that the process of muonization of supernova matter, i.e. the gradual accumulation of a net muon lepton number in the PNS, proceeds in two steps~\cite{Bollig:2017lki,Bollig:2020xdr,Guo:2020tgx,Fischer:2020vie,Capozzi:2020syn}: first, the production of thermal high-energy $\nu_\mu$ and $\bar\nu_\mu$ via neutrino pair processes once temperatures reach $T>10$--$20$~MeV, which secondly, enable muonic weak processes, i.e.~charged current, semi- and purely leptonic reactions. In addition, the presence of muons has been shown to significantly impact the CCSN dynamics~\cite{Bollig:2017lki,Guo:2020tgx}. The additional degrees of freedom soften the equation of state (EOS), accelerating PNS contraction. This in turn enhances the luminosities and mean energies of emitted neutrinos, strengthening post-shock heating and favoring neutrino-driven explosions~\cite{Bollig:2017lki}. 

Muonization arises because more $\nu_\mu$ than $\bar\nu_\mu$ are absorbed by the medium~\cite{Guo:2020tgx}. It is inherently dynamical: shortly after core bounce and during early post-bounce evolution, the abundance of muons is still building up and the muon chemical potential can remain significantly smaller than the electron chemical potential in regions where neutrinos are trapped. Long-term CCSN simulations that follow the PNS deleptonization and cooling phase for several tens of seconds after explosion onset show that, in the PNS interior, the electron and muon chemical potentials equalize only at late times in the PNS interior, on a timescale of several $10$~s (see the Appendix of Ref.~\cite{Fischer:2021jfm}). This timescale depends not only on the hadronic EOS, but also on the macroscopic hydrodynamical evolution. In particular, recent self-consistent multi-dimensional CCSN simulations with spectral neutrino transport show that PNS convection can significantly affect this timescale~\cite{Janka2025ARNPS,Burrows2026arXiv260209025R}.
Below, we make explicit the distinction between the neutrino transport and hydrodynamical timescales governing the macroscopic PNS evolution, and the local chemical-equilibration timescales, set by microscopic weak processes.

Strange hadrons introduce additional degrees of freedom at high density~\cite{1960SvA}, thereby also modifying the EOS with important implications for nuclear astrophysics. Their appearance generally softens the EOS, potentially reducing the maximum neutron star mass below the observed $2 M_\odot$ limit~\cite{Antoniadis13,Cromartie:2020NatAs,NICER_Miller2019,NICER_Watts2019,NICER_Riley2021,NICER_Miller2021} and leading to the so-called ``hyperon puzzle''~\cite{Glendenning1982,Glendenning1985,Vidana:2018bdi}. Proposed resolutions include early quark deconfinement, strong repulsive hyperon interactions, or the inclusion of three-body forces (see Ref.~\cite{Tolos:2020aln} and references therein). Incorporating hyperons in CCSN simulations may therefore offer complementary constraints on their role in dense, hot and isospin asymmetric nuclear matter.

In EOS constructions, strangeness is typically assumed to be in chemical equilibrium, see e.g.~\cite{Oertel:2016bki,Fischer:2024ivh}. Unlike muons, the approach to the local neutron--$\Lambda$ chemical equilibrium condition, $\mu_\Lambda = \mu_n$, is directly controlled by weak-interaction rates (see below) and has been shown to occur on short timescales, of order microseconds, in the hot and dense nuclear matter conditions of neutron star mergers~\cite{Alford:2020pld}. However, a systematic study of these equilibration timescales under CCSN thermodynamic conditions has not yet been carried out. 

In addition, hyperons open new weak-interaction channels that can contribute to neutrino scattering and absorption, potentially modifying neutrino opacities and transport in the PNS. They may also couple to new light particles beyond the Standard Model~\cite{MartinCamalich:2025srw}, which would provide new cooling channels, accelerating PNS deleptonization and shortening the neutrino-emission timescale. Such processes allow CCSN simulations to set stringent bounds on exotic hyperon reactions~\cite{Camalich:2020wac,Cavan-Piton:2024ayu}.

In this paper, we investigate novel $\Lambda$ hyperon production rates in hot, dense, and highly isospin-asymmetric nuclear matter. We focus on $\Lambda$ hyperons because they are the lightest strange baryons and are therefore expected to be among the first strange degrees of freedom populated in dense matter, providing a representative leading contribution to strangeness-related effects.
From these rates, we first estimate the local $\Lambda$ chemical-equilibration timescales under typical PNS conditions, showing that nonleptonic reactions dominate the approach to equilibrium. Second, we compute (anti)neutrino opacities from the semileptonic reactions. This allows us to assess the validity of assuming chemical equilibrium for hyperons in simulations, as well as their potential impact on neutrino transport.

The paper is organized as follows. In Sec.~\ref{sec:SNrates}, we present the framework for local hyperon chemical equilibration, highlighting the differences from electron--muon chemical convergence. We then define the thermal-rate formalism and collision operators, and introduce the representative PNS thermodynamic conditions employed in this work. In Sec.~\ref{sec:weak_hyps}, we compute the nonleptonic weak rates for $\Lambda$ production and show that they control the chemical-equilibration timescale. In Sec.~\ref{sec:SL_transport}, we turn to semileptonic channels, which are subdominant for equilibration but provide additional neutrino and antineutrino opacity channels relevant for transport. Finally, Sec.~\ref{sec:Conclusions} summarizes our main conclusions.

\section{Equilibration of hyperons}
\label{sec:SNrates}
\subsection{Preamble: hyperons vs muons}
\label{sec:preamble}

The build-up of muons and the chemical equilibration of hyperons in CCSNe proceed differently, reflecting the distinct conservation laws and production mechanisms at play.

A key aspect of muon production is the conservation of muon lepton number (neglecting neutrino oscillations), which ties $\mu^\pm$ generation to the absorption of muonic (anti)neutrinos. Muonization proceeds dynamically: $\nu_\mu\bar\nu_\mu$ pairs are created by thermal processes, neutron-rich matter absorbs $\nu_\mu$ more efficiently than $\bar\nu_\mu$, and this asymmetry gradually builds up a net $\mu^-$ population, through semileptonic weak reactions. 
This process is therefore controlled by the coupled evolution of weak-interaction rates and flavor-dependent neutrino transport~\cite{Fischer:2020vie}. Accordingly, the hierarchy $\mu_\mu\ll\mu_e$ in the neutrino-trapped region should not be interpreted, by itself, as a direct measure of local chemical equilibrium. For the semileptonic charged-current reactions on nucleons, local chemical equilibrium requires
\begin{align}
\label{eq:chemical_eq}
\mu_e-\mu_{\nu_e}
=
\mu_\mu-\mu_{\nu_\mu}
=
\mu_n-\mu_p .
\end{align}
In the neutrino-transparent limit relevant for cold neutron stars, where the relevant neutrino chemical potentials vanish, the condition above reduces to $\mu_e = \mu_\mu$. A similar convergence can already occur during the post-bounce phase near the neutrinosphere, where neutrinos start to decouple from PNS matter. Thus, we use ``\textit{local chemical equilibrium}'' to denote the condition in Eq.~\eqref{eq:chemical_eq} imposed by the relevant weak reactions, and distinguish it from the late-time ``\textit{electron--muon chemical convergence}'', $\mu_e \simeq \mu_\mu$, driven by flavor-dependent transport during PNS deleptonization.

On the other hand, the chemical equilibration of hyperons proceeds in a qualitatively different way. 
In the Standard Model, strangeness is not conserved by weak interactions: hyperons, such as the $\Lambda$, can be produced directly either through semileptonic or nonleptonic reactions (the latter being the dominant ones, as will be demonstrated below). 
Their production does not depend on fluxes of lepton flavor, and their abundances are determined by the balance of local weak reaction rates and the evolving thermodynamic conditions. Thus, the relaxation toward $\Lambda$ chemical equilibrium, i.e.~$\mu_\Lambda= \mu_n$, occurs on a timescale set by the fastest $\Lambda$-changing weak processes, rather than by the evolution driven by neutrino transport that controls the late-time electron--muon chemical convergence.
The pertinence of the equilibrium approximation then relies on the comparison of this timescale with the shortest timescales over which the medium changes its properties
e.g. the hydrodynamical and convective timescales, and the neutrino-transport timescales associated with PNS deleptonization and cooling.

\subsection{Thermal rates}
\label{sec:thermal_rates}

We adopt the standard kinetic-theory definition of the thermal absorption rate of particle $a$ in the medium for processes of the form $a + b + c \ldots \longrightarrow  i + j + k \ldots$:
\begin{align}
\label{eq:Rategen}
\Gamma_a(E_a)&= \frac{1}{2E_a\mathfrak{g}_a}\int \Big[\prod {\rm d} \Pi_\alpha f_\alpha \Big] \Big[ \prod {\rm d}\Phi_\omega (1\pm f_\omega)\Big] \nonumber\\
&\times (2\pi)^4 \delta^4 (p_a+\sum p_\alpha-\sum p_\omega) \sum_{\rm spins}|\mathcal{M}|^2 \,.
\end{align}
These integrals involve $|\mathcal{M}|^2$, the squared matrix element of the process, integrated over the phase space, ${\rm d}\Pi={\rm d}\vec p^{\;3}/(2\pi)^3/2E$, of the initial particles $\alpha=b,c\ldots$ and final particles $\omega=i,j,k,\ldots$. The integrand is weighted by the number density functions $f_\alpha$, together with the Pauli blocking or Bose enhancement factors $(1 \mp f_\omega)$ and $\mathfrak g_a$ is the number of degrees of freedom of the particle $a$. 

The absorption rate for a given process can be expressed in terms of an energy-dependent mean free path of particle $a$ in the medium, $\lambda_a(E_a)$, or equivalently in terms of the corresponding opacity, $\chi_a(E_a)$:
\begin{align}
\label{eq:mfp}
\lambda_a(E_a) = \frac{v_a}{\Gamma_a(E_a)} = \frac{1}{\chi_a(E_a)} ,
\end{align}
where $v_a = |\vec p_a| / E_a$ is the particle velocity.
In general, for transport phenomena, the angular dependence of the opacities must be retained (for details, see Ref.~\cite{Fischer12}). 
Note that the quantities in Eq.~\eqref{eq:mfp} are partial quantities, i.e.~obtained from the partial rate of a specific process.
The total mean free path is instead
\begin{align}
\label{amend}
\lambda_a^{\rm tot}(E_a) = \frac{v_a}{\Gamma_a^{\rm tot}(E_a)}~,
\qquad
\Gamma_a^{\rm tot}(E_a)=\sum_b \Gamma_{ab}(E_a)~,
\end{align}
where $b$ denotes all viable targets for particle $a$ to scatter against or be absorbed by. In the following, we use the partial definition in Eq.~\eqref{eq:mfp} to allow for direct comparison with the literature \cite{Guo:2020tgx,Fischer:2020vie}.

At the level of number densities, the Boltzmann equation for the number density $n_a$ of species $a$ reads
\begin{align}
\label{eq:BoltzNa}
\frac{\mathrm d n_a}{\mathrm d t}=\mathcal C_{\rm prod}-\mathcal C_{\rm abs}~,
\end{align}
where $\mathcal C_{\rm prod}$ and $\mathcal C_{\rm abs}$ are the production and absorption collision operators, respectively, describing the rate of events per unit volume. For a given process, the absorption collision operator is related to the corresponding absorption rate through
\begin{align}
\label{eq:collOp}
\mathcal C_{\rm abs}=\mathfrak{g}_a\int \frac{{\rm d}^3 \vec p_a}{(2\pi)^3}f_a\Gamma_a(E_a)\,.
\end{align}
At equilibrium, the following relation is fulfilled: 
\begin{align}
\label{eq:detailed}
\mathcal C_{\rm abs}=\mathcal C_{\rm prod}~.
\end{align}

\subsection{Thermodynamical conditions}
The relevance of hyperons in CCSN depends sensitively on the EOS assumed for hot and dense nuclear matter.
At present, the high-density EOS remains an active subject of research in nuclear astrophysics and one of the major uncertainties in CCSN simulations~\cite{Janka2025ARNPS,Burrows2026arXiv260209025R}. 
A variety of EOS models have been proposed to describe matter under CCSN conditions, including non-relativistic Skyrme functional models~\cite{LSEOS}, nuclear relativistic mean field models~\cite{Shen98,gshen2011b,HS}, models that are linked to neutron star observations and radius determinations~\cite{SFH}, as well as EOSs including a phase transition to deconfined quark matter~\cite{Sagert09,Klaehn:2017,Bastian:2021} and strange hadrons~\cite{Kochankovski:2023trc}. Our study aims to motivate more self-consistent simulations, where the feedback of hyperons onto PNS conditions is taken into account and quantitatively assessed. 

For the thermodynamical conditions at the core of the PNS we use the conditions obtained in the simulation reported in Ref.~\cite{Guo:2020tgx},  for a 20 $M_\odot$ progenitor at 0.4~s postbounce and at a radius of $r\simeq13.6$ km (referred to as ``condition A'' in that reference):
\begin{equation}
\boxed{
\arraycolsep=1.4pt\def\arraystretch{1.5}
\begin{array}{c}
\text{ \bf PNS conditions}
\\
\;\;\;T = 38.3 \, \MeV,\;\rho=10^{14}$ g cm$^{-3},\;\;\;\\
\;\;\;\mu_\mu =64.1 \, \MeV,\; \mu_{\nu_\mu} =-20  \, \MeV,\;\;\;\\
\;\;\;\mu_e = 83.3 \, \MeV,\; \mu_{\nu_e} =-2.1\, \MeV,\;\;\;\\ 
\;\;\;\mu_n =886.0 \, \MeV,\; U_n =-24.9 \, \MeV,\;\;\;\\
\;\;\;\mu_p =800.7 \, \MeV,\; U_p =-42.9 \, \MeV.\;\;\;\\
\end{array}
}
\label{eq:typicalPNS}
\end{equation}
These quantities are predicted by the Lattimer-Swesty EOS~\cite{LSEOS} with compressibility modulus of $220$~MeV (LS220), which was employed in the simulations and includes the chemical potentials of the different species and the nucleon single-particle potentials. In the next section, we will explain how we extended these conditions to include $\Lambda$-hyperons.
The baryonic mean-field potentials $U$ enter the in-medium dispersion relation 
for which we employ here the relativistic version for convenience,
\begin{align}
\label{eq:disprel}
E=\sqrt{\vec p^2+M_{\rm eff}^2}+U,
\end{align}
where $M_{\rm eff}$ is the effective baryon mass, which in the Lattimer--Swesty EOS is set equal to its physical value, $M_{\rm eff}=M$.

The conditions in Eq.~\eqref{eq:typicalPNS} imply that the electron and muon weak reactions on nucleons are approximately in local chemical equilibrium, since Eq.~\eqref{eq:chemical_eq} is nearly satisfied. 
At the same time, $\mu_e\neq\mu_\mu$, indicating that the late-time electron--muon convergence characteristic of the neutrino-transparent limit has not yet been reached.

\section{Nonleptonic weak rates and hyperon equilibration}
\label{sec:weak_hyps}

Several weak strangeness-changing processes contribute to hyperon equilibration in nuclear matter. On the one hand, there are \textit{nonleptonic reactions} induced by the charged-current quark transitions $u + d \leftrightarrows s + u$. We classify them into nonleptonic scattering processes,
\begin{align}
\label{eq:HADscat}
    B + B  &\leftrightarrows  B^\prime + B,
\end{align}
and nonleptonic coalescence and decay processes,
\begin{align}
\label{eq:HADcoal}
\pi + B \leftrightarrows B^\prime.
\end{align}
The latter are induced by thermal pions that may emerge in dense nuclear matter~\cite{PhysRevLett.29.382,PhysRevLett.30.1340,Fore:2019wib}. Here $B$ and $B^\prime$ denote octet baryons connected by a weak transition with strangeness change $\Delta S=-1$ (e.g. $B^\prime=\Lambda$, $B=n$). For the calculation of the nonleptonic processes, we will adopt an effective field theory (EFT) framework specifically tailored to describe the weak $\Lambda + N \longrightarrow N+N$ transition, developed in analyses of nonmesonic hypernuclear decays~\cite{Parreno:2003ny,Parreno:2003nx,Perez-Obiol:2013waa}. This framework already includes, as an ingredient, the weak $\Lambda N\pi$ vertices that describe the nonleptonic coalescence process.

As we show in the next subsection, these nonleptonic channels dominate the equilibration dynamics and drive the $\Lambda$-hyperon population toward chemical equilibrium. Semileptonic reactions can also contribute to the overall hyperon equilibration rate in the PNS. As shown in Sec.~\ref{sec:SL_transport}, however, this contribution is negligible compared with the nonleptonic one, while they may play a role in neutrino transport dynamics in CCSNe.

For the estimate of the $\Lambda$-hyperon equilibration time, we take the PNS thermodynamic conditions and composition of Eq.~\eqref{eq:typicalPNS} as a fixed reference background and augment it with a population of $\Lambda$ hyperons. The corresponding equilibrium population is determined by imposing the chemical-equilibrium condition $\mu_\Lambda=\mu_n$. For the potential $U_\Lambda$,  we use the hyperonic extension of this LS220, called LS220$\Lambda$, through the {\tt CompOSE} interpolation tables, obtaining  $U_\Lambda=-16.7$ MeV.\footnote{We note that the hyperonic version of the LS220 EOS overproduces strangeness at $T=0$ and is unable to obtain $2M_\odot$ neutron stars. However, at the high temperatures characteristic of CCSN it predicts $\Lambda$ abundances equivalent to other hyperonic EOS.}

This procedure should be understood as a minimal extension of the non-hyperonic
background: the thermodynamic state and the abundances of the non-hyperonic
species are kept fixed when the $\Lambda$ component is added. This approximation is justified a posteriori by the small $\Lambda$ population found at equilibrium  $n_\Lambda^{\rm eq}/n_n^{\rm eq}\sim 10^{-2}$, consistent with the corresponding dilute phase-space occupation $f_\Lambda^{\rm eq}\ll 1.$\footnote{
For the equilibrium conditions used here, the maximum occupation is
$f_\Lambda\lesssim {\rm few}\times 10^{-3}$.
}

We then study the local relaxation of the $\Lambda$ abundance on top of this fixed background. In the dilute limit, final-state Pauli blocking of produced $\Lambda$ hyperons is negligible, so that $\mathcal C_{{\rm prod},\Lambda}$ can be treated as approximately independent of $n_\Lambda$ during the relaxation. Writing
\begin{align}
    n_\Lambda = n_\Lambda^{\rm eq}+\delta n_\Lambda ,
\end{align}
and expanding Eq.~\eqref{eq:BoltzNa} around equilibrium gives
\begin{align}
    \frac{\mathrm d \delta n_\Lambda}{\mathrm d t}
    =
    \left(
    \left.
    \frac{\partial \mathcal C_{{\rm prod},\Lambda}}{\partial n_\Lambda}
    \right|_{\rm eq}
    -
    \left.
    \frac{\partial \mathcal C_{{\rm abs},\Lambda}}{\partial n_\Lambda}
    \right|_{\rm eq}
    \right)\delta n_\Lambda
    +\mathcal O(\delta n_\Lambda^2),
\end{align}
where the zeroth-order term vanishes at equilibrium by Eq.~\eqref{eq:detailed}. Neglecting
$\partial\mathcal C_{{\rm prod},\Lambda}/\partial n_\Lambda$ under the assumptions
above, one obtains
\begin{align}
\label{eq:lambda_linearized}
    \frac{\mathrm d \delta n_\Lambda}{\mathrm d t}
    =
    -\left.
    \frac{\partial \mathcal C_{{\rm abs},\Lambda}}
    {\partial n_\Lambda}
    \right|_{\rm eq}
    \delta n_\Lambda .
\end{align}
This defines the local equilibration time
\begin{align}
\label{eq:tau_lambda_def}
    \tau_\Lambda^{-1}
    =
    \left.
    \frac{\partial \mathcal C_{{\rm abs},\Lambda}}
    {\partial n_\Lambda}
    \right|_{\rm eq}.
\end{align}
It remains to estimate the derivative in Eq.~\eqref{eq:tau_lambda_def}. At fixed
background, variations of $n_\Lambda$ can be parametrized by variations of
$\mu_\Lambda$. Using Eq.~\eqref{eq:collOp}, one finds
\begin{align}
    \left.
    \frac{\partial \mathcal C_{{\rm abs},\Lambda}}
    {\partial n_\Lambda}
    \right|_{\rm eq}
    =
    \frac{
    \displaystyle
    \int \dfrac{\mathrm d^3 p_\Lambda}{(2\pi)^3}
    \left.
    \frac{\partial f_\Lambda}{\partial \mu_\Lambda}
    \right|_{\rm eq}
    \Gamma_\Lambda(E_\Lambda)
    }
    {
    \displaystyle
    \int \dfrac{\mathrm d^3 p_\Lambda}{(2\pi)^3}
    \left.
    \frac{\partial f_\Lambda}{\partial \mu_\Lambda}
    \right|_{\rm eq}
    } .
\end{align}
In the dilute limit $\left.\partial f_\Lambda/\partial \mu_\Lambda \right|_{\rm eq} \simeq f_\Lambda^{\rm eq}/T$, so that the absorption collision operator is approximately linear in $n_\Lambda$, and
\begin{align}
\label{eq:tau_lambda_approx}
    \tau_\Lambda^{-1}
    \simeq
    \frac{
    \displaystyle
    \int \dfrac{\mathrm d^3 p_\Lambda}{(2\pi)^3}
    f_\Lambda^{\rm eq}\,
    \Gamma_\Lambda(E_\Lambda)
    }
    {
    \displaystyle
    \int \dfrac{\mathrm d^3 p_\Lambda}{(2\pi)^3}
    f_\Lambda^{\rm eq}
    }
    =
    \frac{\mathcal C_{{\rm abs},\Lambda}^{\rm eq}}
    {n_\Lambda^{\rm eq}}
    \equiv
    \bar\Gamma_\Lambda^{\rm eq}.
\end{align}

Therefore, the equilibrium conditions specified above, including the $\Lambda$ component, are sufficient to estimate the local equilibration timescale of $\Lambda$ hyperons. Throughout the paper, we use Eq.~\eqref{eq:tau_lambda_approx} as a common measure of the equilibration time associated with each nonleptonic and semileptonic process considered below.

\subsection{Nonleptonic scattering}
\label{sec:HADscat}

We now estimate $\Delta S=1$ strangeness production in nonleptonic baryon-baryon scattering using the EFT framework for the 
$\Lambda + N \longrightarrow N+ N$ 
weak transitions~\cite{Parreno:2003ny,Parreno:2003nx,Perez-Obiol:2013waa}. In particular, we retain the leading-order (LO) contributions, consisting of 
long-range one-meson-exchange diagrams, mediated by pions and kaons, together 
with short-range contact interactions~\cite{Parreno:2003ny}.
In this section, we focus on the two nonleptonic scattering channels relevant for $\Lambda$ production, namely $n+n \longrightarrow  n+\Lambda$ and $p+n\longrightarrow p+\Lambda$, and study them within this EFT framework.

\subsubsection{One meson exchange}
\label{sec:onemeson}

\begin{figure*}
\begin{tabular}{cc}
    \centering
    \includegraphics[width=0.45\linewidth]{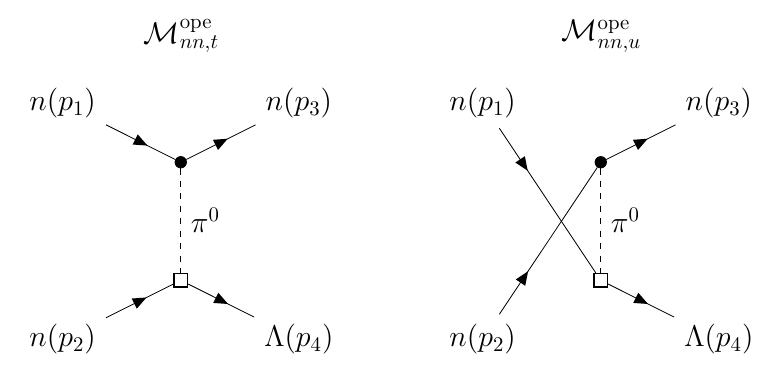}
   & \includegraphics[width=0.45\linewidth]{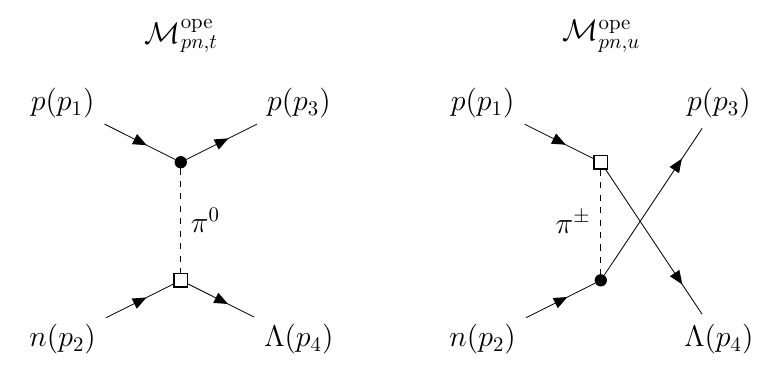}
\end{tabular}
 \caption{One--pion--exchange contributions to $n+n\longrightarrow n+\Lambda$ (two diagrams to the left) and $p+n\longrightarrow p+\Lambda$ (two diagrams to the right): in both cases we show $t$-channel and crossed $u$-channel diagrams. The weak vertex is indicated by an open square, while the strong vertex is denoted by a filled circle.}
    \label{fig:OPE}
\end{figure*}

\begin{figure*}
\begin{tabular}{cc}
    \centering
    \includegraphics[width=0.45\linewidth]{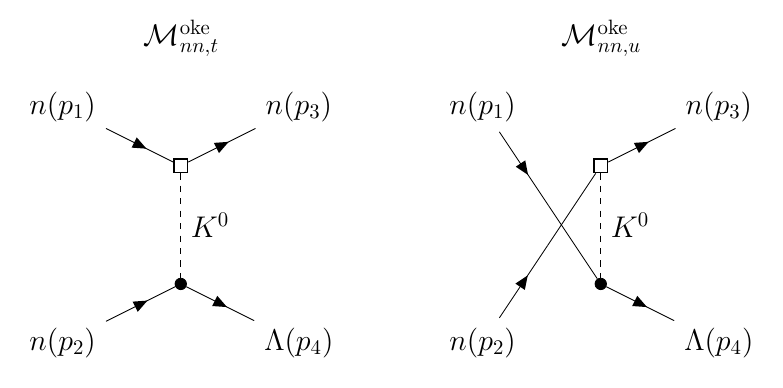}
   & \includegraphics[width=0.45\linewidth]{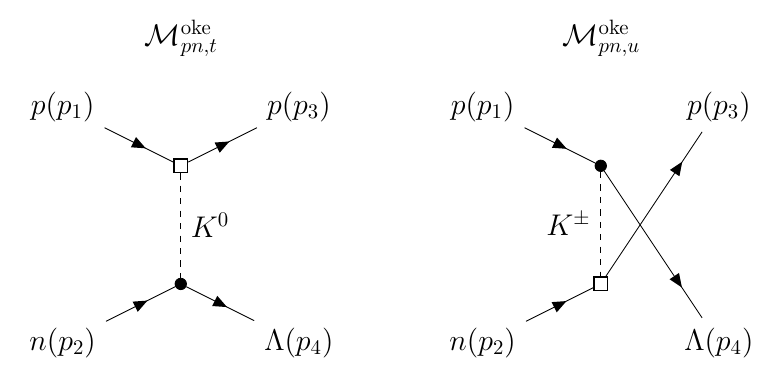}
\end{tabular}
 \caption{One--kaon--exchange contributions with the same conventions as in Fig.~\ref{fig:OPE}.
 }
    \label{fig:OKE}
\end{figure*}

The one-meson-exchange contributions arise from diagrams in which a pion or 
kaon is exchanged between two baryon lines, with one weak vertex inducing the 
$\Delta S=1$ transition and one strong vertex, see Figs.~\ref{fig:OPE} and \ref{fig:OKE}. The relevant nonleptonic weak  vertices, describing  $B B'\phi$-interactions, where $\phi$ denotes a pion or kaon, are written in the chiral-Lagrangian convention as
\begin{align}
\label{eq:LagWeak}
\mathcal L_W^{\rm ome} \supset
i\,
G_F m_{\pi^+}^2\,
\bar B'\left(
\mathcal A_{\phi B'B}
-\gamma_5\,\mathcal B_{\phi B'B}
\right)B\,\phi^\dagger\,+\,\mathrm{h.c.}~,
\end{align}
where $G_F m_{\pi^+}^2 = 2.2720935\times 10^{-7}$, 
and $\mathcal A_{\phi B'B}$ and $\mathcal B_{\phi B'B}$ are the dimensionless
parity-violating (PV) and parity-conserving (PC) amplitudes, respectively, in 
the standard convention used for nonleptonic hyperon decays~\cite{ParticleDataGroup:2024cfk}. Note that with this phase convention $\mathcal A_{\phi^\dagger B B^\prime}=-\mathcal A_{\phi  B^\prime B}$ from the hermitian conjugate.
The values of the amplitudes used in the calculations are obtained from the nonleptonic hyperon decay rates and asymmetries in vacuum, which are listed in App.~\ref{app:AB_extraction}. This interaction also determines the pionic coalescence process $\pi + N \longrightarrow \Lambda$ (see below Sec.~\ref{sec:HADcoal}), which is the inverse of the nonleptonic $\Lambda$ weak decay.

The strong interaction vertex is described by the standard LO chiral Lagrangian~\cite{Pich:1995bw},
\begin{equation}
\label{eq:LagStrong}
\mathcal{L}_S^{\rm ome}\supset-\frac{\mathcal C_{\phi B^\prime B}}{2f_\pi}\;
\bar B^\prime\gamma^\mu\gamma_5B\,\partial_\mu\phi^\dagger\,,
\end{equation}
where $\mathcal{C}_{\phi B^\prime B}$ are dimensionless constants that depend on the two baryon couplings $D=0.80$ and $F=0.46$, where $g_A=D+F$ is the axial nucleon coupling~\cite{Cabibbo:2003cu}, and $f_\pi=92.4$ MeV is the pion decay constant~\cite{Pich:1995bw}.\footnote{We adopt an overall minus sign in Eq.~\eqref{eq:LagWeak} relative to Refs.~\cite{Parreno:2003ny,Parreno:2003nx} so that the one-meson-exchange amplitudes retain the same signs. This compensates for the fact that the chiral strong Lagrangians are on-shell equivalent to the pseudoscalar ones employed in those references, but differ by an overall minus sign~\cite{Scherer:2012xha}.}  

The one--pion--exchange (OPE) contribution to 
$n + n \longrightarrow n + \Lambda$ proceeds via $\pi^0$ exchange. 
Because the two incoming neutrons are identical fermions, 
two topologically distinct diagrams contribute, corresponding to the two possible attachments of the exchanged pion to the external neutron legs. 
We denote the corresponding amplitudes by 
$\mathcal{M}^{\rm ope}_{nn,t}$ and $\mathcal{M}^{\rm ope}_{nn,u}$, such that
\begin{equation}
\label{eq:M_nn}
    \mathcal{M}_{nn}^{\rm ope}
    = \mathcal{M}^{\rm ope}_{nn,t} - \mathcal{M}^{\rm ope}_{nn,u} \,.
\end{equation}
The relative minus sign arises from the exchange of two external fermion lines and hence from Fermi statistics (equivalently, from Wick's theorem). In case of the channel $p + n \longrightarrow p + \Lambda$, there are two contributions involving either $\pi^0$ or $\pi^{\pm}$ exchange depending on which of the initial nucleons converts into a $\Lambda$ through the weak vertex. We denote the corresponding amplitudes by $\mathcal M^{\rm ope}_{pn,t}$ and $\mathcal M^{\rm ope}_{pn,u}$, so that
\begin{equation}
\label{eq:M_pn}
    \mathcal M_{pn}^{\rm ope}=\mathcal M^{\rm ope}_{pn,t}+\mathcal M^{\rm ope}_{pn,u}\,.
\end{equation}
with the sign of each term fixed by the corresponding charge-basis vertices and Wick contractions. 

The one--kaon--exchange (OKE) contribution is organized in close analogy with the OPE one.
The important difference is that the exchanged kaon carries strangeness, so the
$N\longleftrightarrow\Lambda$ conversion now takes place at the strong derivative
vertex in Eq.~\eqref{eq:LagStrong}, while the weak vertex couples the kaon to
nucleons. For $n + n \longrightarrow n + \Lambda$, this implies $K^0$ exchange
only, and the two contractions are
antisymmetrized,
\begin{equation}
\mathcal M_{nn}^{\rm oke}
=
\mathcal M^{\rm oke}_{nn,t}
-
\mathcal M^{\rm oke}_{nn,u}\, .
\end{equation}
For $p + n \longrightarrow p +\Lambda$, charged-kaon exchange can also contribute and the two topologies are summed,
\begin{equation}
\mathcal M_{pn}^{\rm oke}
=
\mathcal M^{\rm oke}_{pn,t}
+
\mathcal M^{\rm oke}_{pn,u}\, .
\end{equation}

We present the relevant diagrams in Figs.~\ref{fig:OPE} and~\ref{fig:OKE} and carry out the calculation in a fully relativistic framework (see App.~\ref{app:SL_sqme}).

\subsubsection{Contact interactions}
\label{sec:contact_interactions}
The contact interactions in the EFT are parametrized in terms of the non-relativistic $\Lambda + N \longrightarrow N + N$ potential. 
It is therefore useful to first write the corresponding operators in a nonrelativistic heavy-baryon
language. For the $p+n \longrightarrow p+\Lambda$ channel, after performing a Fierz rearrangement (for details, see App.~\ref{app:CTs}), the LO contact operator can be
written in terms of two couplings as
\begin{align}
\label{eq:CTs}
\mathcal L_W^{\rm ct} \supset- G_F&\Big[C_S^{pn}(\Lambda^\dagger n)(p^\dagger p)\nonumber\\
&+C_T^{pn}(\Lambda^\dagger\vec\sigma n)\cdot(p^\dagger\vec\sigma p)\Big]+{\rm h.c.}
\end{align}
where $p$, $n$, and $\Lambda$ are two-component heavy-baryon fields. 
The $n + n \longrightarrow n + \Lambda$ channel is described by the analogous operator obtained by replacing $p\to n$ and $C_{S,T}^{pn}\to C_{S,T}^{nn}$ in Eq.~\eqref{eq:CTs}. These are low-energy constants (LECs) whose values can be determined from data or in models. In our analysis, we will use the values obtained from the fit to hypernuclear data in Ref.~\cite{Parreno:2003ny},
\begin{align}
\label{eq:LECsdata}
&C_S^{nn}=-9.95,&C_T^{nn}=-5.62,\nonumber\\
&C_S^{pn}=0.54,&C_T^{pn}=-2.13.
\end{align}

Equation~\eqref{eq:CTs} defines the contact interaction at the nonrelativistic level. In the evaluation of the matrix elements, we promote these structures to a minimal covariant embedding, chosen such
that its leading nonrelativistic reduction reproduces Eq.~\eqref{eq:CTs}. This
allows for the contact amplitude to be added coherently to the relativistic OPE and
OKE amplitudes. The explicit embedding and the associated ambiguity are discussed in App.~\ref{app:CTs}.

\begin{figure*}
\centering
\begin{tabular}{c}
    \includegraphics[width=0.7\linewidth]{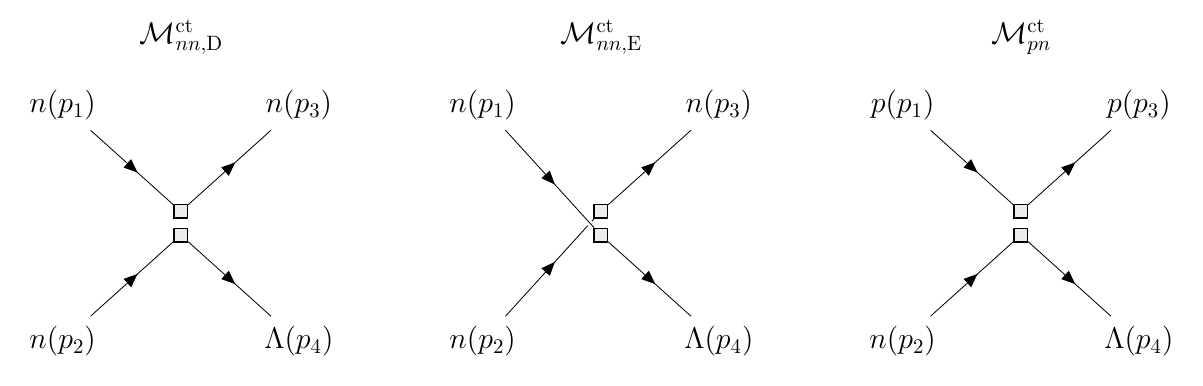}
\end{tabular}
 \caption{Contact-term contributions to $n + n \longrightarrow n + \Lambda$ (left panel) and $p + n \longrightarrow p + \Lambda$ (right panel). The gray squares indicate which fermion legs belong to the same baryon bilinear in the four-baryon contact operator.}
    \label{fig:CT}
\end{figure*}

As in the one--meson--exchange cases, there are two distinct diagrams for $n + n \longrightarrow n + \Lambda$, due to the presence of two identical neutrons in the initial state. We label the corresponding direct and exchange contributions by D and E, respectively, and write,
\begin{equation}
\mathcal M^{\rm ct}_{nn}
=
\mathcal M^{\rm ct}_{nn,\rm{D}}-\mathcal M^{\rm ct}_{nn,\rm{E}}.
\end{equation}
On the other hand, for $p+n\longrightarrow p+\Lambda$, the contact term is represented by a single proton-spectator amplitude $\mathcal M^{\rm ct}_{pn}$, in which the neutron is converted into a $\Lambda$ while the proton line remains a proton. Fig.~\ref{fig:CT}  shows the contact term contributions considered in the two channels.

This EFT description can be compared with the model often adopted in the literature~\cite{Alford:2020pld}, where the transition is described as a charge-exchange process mediated by a $W^-$ boson between the $\Lambda p$ and $p n$ vertices, with the corresponding matrix element factorized into separate hadronic currents. In this approximation, one obtains~\footnote{We have included in this calculation the vector $\Lambda p$ coupling $f_1^{\Lambda p}(0)=-\sqrt{3/2}$, which was omitted in previous literature, see App.~\ref{app:CTs}.}
\begin{align}
\label{eq:LECsmodel}
&C_S^{W,pn}=0.17,&C_T^{W,pn}=-0.18,
\end{align}
while the process $n + n \longrightarrow n + \Lambda$ receives no contribution. The EFT predictions differ significantly from those relying on this $W$-boson-exchange model: for the $p + n \longrightarrow  p + \Lambda$ channel, the corresponding collision operators differ by about one order of magnitude under the same PNS conditions of Eq.~\eqref{eq:typicalPNS}.

Several caveats apply to the use of EFT contact terms in this context. First, the LECs have been extracted with a specific Gaussian regulator, although its impact on the thermal rates is subdominant compared to other uncertainties.\footnote{The regulator in momentum space corresponds to a form factor $f(-\vec q^2/2m_\rho^2)$, while high-momentum contributions are already suppressed by the distribution tails. We find that including this form factor reduces the contact-term contribution by $4\%$.} More importantly, the limited and imprecise hypernuclear decay data used in the fits lead to sizable uncertainties in the LECs. In addition, these LECs are determined in finite hypernuclei, where nuclear structure and final-state effects are modeled, and are here extrapolated to hot and dense CCSN matter, where many-body and thermal effects are more important.
 
\subsubsection{Calculation of the nonleptonic scattering rates}
\label{eq:hadronic calculations}

Our goal is to compute the scattering rates entering the collision operators for these two processes. To this end, we now introduce the approximations used in the rate calculation for these scattering channels.

\begin{itemize}
\item We approximate the contribution of the Pauli blocking factors $(1\pm f_\omega)$ in the integrand of Eq.~\eqref{eq:Rategen} by their thermal averages,
\begin{align}
\label{eq:Fdeg}
F_\omega=\frac{\mathfrak g_\omega}{n_\omega}\int \frac{{\rm d}^3\vec p_\omega}{(2\pi)^3} f_\omega(1-f_\omega).
\end{align}
\item We use the relativistic dispersion relation for the baryons 
\begin{align}
\label{eq:ApproxDispRel}
    E=\sqrt{M_*^2+\vec p^{\,2}},
\end{align}
where $M_*=M+U$. This is a good approximation provided that $U/M\ll 1$ (which is the case for the chosen EOS) and allows us to simplify the phase space integrals.
\end{itemize}
These approximations are adopted for simplicity and provide sufficient accuracy for the purposes of this work. They do not fulfill exactly the condition in Eq.~\eqref{eq:detailed}; nevertheless, under the PNS conditions of Eq.~\eqref{eq:typicalPNS}, the residual imbalance in the nonleptonic scattering collision operators remains at the level of $\Delta \mathcal{C}/\mathcal{C}_{\rm tot}\sim 15\%$.

In this simplified setup, the absorption width $\Gamma_a$ for a generic scattering process $a + b \longrightarrow i + j$ can be written in terms of the scattering cross section $\sigma(s)$ as
\begin{align}
\label{eq:RateScat}
\Gamma_a(E_a)=\frac{\mathfrak g_b F_i F_j}{8\pi^2 E_a |\vec p_a|}\int^\infty_{E_0}{\rm d}E_bf_b\int^{s_{\rm max}}_{s_0}{\rm d} s\sqrt{s}p\,\sigma(s),
\end{align}
where $s_0$ is the reaction threshold, $s_{\rm max}$ is the maximum center-of-mass energy for given $E_a$ and $E_b$, and $p=\lambda_{\rm K}(s^2,m_a^2,m_b^2)^{1/2}/2\sqrt{s}$ is the center-of-mass momentum, with $\lambda_{\rm K}$ denoting the K\"{a}ll\'en function.
After integrating $\Gamma_a$ as in Eq.~\eqref{eq:collOp}, we use Eq.~\eqref{eq:tau_lambda_approx} to infer the $\Lambda$-hyperon equilibration timescales for both scattering processes, reported in Tab.~\ref{tab:HADtimescales}.

Under the PNS conditions of Eq.~\eqref{eq:typicalPNS}, the collision operator for both the $n + n \longrightarrow  n + \Lambda$ and $p + n \longrightarrow p+ \Lambda$ channels is almost entirely dominated by the contact interaction, while the OPE and OKE contributions yield only negligible and comparable corrections. This contrasts with earlier estimates at lower temperatures, where the OPE contribution was found to exceed the $W$-boson-exchange contact term by up to three orders of magnitude, see Fig.~3 in \cite{Alford:2020pld}. To assess this model dependence, we repeat the calculation under the same PNS conditions using the smaller $W$-boson-exchange contact terms of Eq.~\eqref{eq:LECsmodel} instead of the EFT values of Eq.~\eqref{eq:LECsdata}. The resulting total $p + n \longrightarrow p + \Lambda$ collision operator is reduced by about one order of magnitude relative to the EFT prediction.

\subsection{Nonleptonic coalescence} 
\label{sec:HADcoal}

We now focus on the nonleptonic coalescence channels in Eq.~\eqref{eq:HADcoal} that directly produce $\Lambda$ hyperons in the medium.
\begin{align}
p + \pi^- \longrightarrow \Lambda~,
\qquad
n + \pi^0 \longrightarrow \Lambda~.
\label{eq:HADcoal_channels_explicit}
\end{align}
The corresponding weak interaction vertices are described by the effective Lagrangian in Eq.~\eqref{eq:LagWeak}. Using Eq.~\eqref{eq:collOp} together with two-body decay kinematics, we can express the coalescence collision operator as
\begin{align}
\mathcal C_{\rm prod}^\pi=\frac{M_\Lambda^2\Gamma_\pi}{2\pi^2|\vec k|}\int^\infty_{m_\pi} {\rm d}E_\pi f_\pi\int^{E_+}_{E_-}{\rm d}E_N f_N(1-f_\Lambda),
\end{align}
where 
\begin{align}
&E_{\pm}=\frac{1}{2m_\pi^2}\left((M_\Lambda^2-M_N^2-m_\pi^2)E_\pi\pm2M_\Lambda\sqrt{E_\pi^2-m_\pi^2}\,|\vec k|\right),\nonumber\\
&|\vec k|=\frac{1}{2M_\Lambda}\lambda_{\rm K}^{1/2}(M_\Lambda^2,M_N^2,m_\pi^2),
\end{align}
$\Gamma_\pi$ is the vacuum $\Lambda$ decay width of the given channel (into $p+ \pi^-$ or $n+ \pi^0$) and each of the distribution functions depends on the energy (or 3-momentum of the corresponding particle). 

We then compute the equilibration timescales from the charged and neutral channels separately, following Eq.~\eqref{eq:tau_lambda_approx}. Here we make the additional assumption that pions are in chemical equilibrium with the surrounding hadronic medium via strong processes. Hence, we take the chemical potentials of pions fixed by conserved baryon number and electric charge, $\mu_{\pi^0}=0$ and $\mu_{\pi^-}=\mu_n-\mu_p$. 

The resulting nonleptonic equilibration timescales are summarized in Tab.~\ref{tab:HADtimescales}. At this PNS condition, $\Lambda$ hyperons equilibrate on a very short timescale, of order $10^{-2}\,\mathrm{ns}$. The equilibration is dominated by nonleptonic scattering, especially the $n + n \longrightarrow n + \Lambda$ channel, whose rate is about a factor of six
larger than that of $p + n \longrightarrow p+ \Lambda$. This hierarchy can be traced to the relative values of the fitted contact LECs in Eq.~\eqref{eq:LECsdata}. Nonleptonic coalescence channels are slower than the nonleptonic scattering rates by more than an order of magnitude, with characteristic timescales of order $0.5$--$0.8\,\mathrm{ns}$ at equilibrium conditions $\mu_{\pi^0}=0$ and $\mu_{\pi^-}=\mu_n-\mu_p$. 

The nonleptonic $\Lambda$ chemical-equilibration time is several orders of magnitude shorter than the characteristic macroscopic timescales over which the PNS background evolves. 
These range from the hydrodynamical or free-fall timescale $\sim10^{-3}$~s for a characteristic density $\bar\rho\simeq10^{14}$~g~cm$^{-3}$~\cite{BurrowsLattimes:986ApJ307,Nagakura:2020mnras}, to convective timescales of order $10^{-2}$--$10^{-1}$~s~\cite{Mezzacappa:1998ApJ493,Dessart:2006ApJ645}, and to the longer PNS deleptonization and Kelvin--Helmholtz cooling timescales, $\sim1$--$10$~s and $\sim10$--$60$~s, respectively~\cite{Fischer:2024PrPNP13704107F}. 
Thus, for the conditions considered here, local $\Lambda$ chemical equilibrium is reached effectively instantaneously on macroscopic PNS timescales.

We also checked the sensitivity of the coalescence rates to simple variations
of the pion sector. Moderate variations of
$\mu_{\pi^-}$ or of the negative pion dispersion relation, through shifts in $E_\pi-$ induced by a pion self-energy down to $\Sigma_{\pi^-}=-60~{\rm MeV}$ as in Fig.~6 of \cite{Fore:2019wib}, do not change the hierarchy. The only cases in which $\Lambda$
production from inverse decay becomes comparable to the scattering contribution
occur when the neutral pion population is driven close to Bose saturation,
i.e.
\begin{align}
    E_{\pi^0,\min}-\mu_{\pi^0}  \lesssim \mathcal{O}(1\text{--}10)\,\mathrm{MeV}.
\end{align}
Such configurations are not expected if neutral pions remain chemically
equilibrated by strong processes, for which $\mu_{\pi^0}=0$. Therefore, under
pion chemical equilibrium, nonleptonic decay/coalescence provides a subleading
contribution to $\Lambda$ equilibration compared with nonleptonic 
scattering.

\begin{table}
\renewcommand{\arraystretch}{1.7}
\setlength{\arrayrulewidth}{.25mm}
  \setlength{\tabcolsep}{0.5 em}
    \centering
    \begin{tabular}{|c|c|}
    \hline
    Reaction     & $\tau_\Lambda$ ($10^{-9}$ s) \\
    \hline
    $n + n \longrightarrow n + \Lambda$  & $0.01$ \\
    \hline
   $p + n \longrightarrow p + \Lambda$ & 0.06 \\
   \hline
   $p + \pi^- \longrightarrow \Lambda$ & 0.5 \\
   \hline
   $n + \pi^0 \longrightarrow \Lambda$ & 0.8 \\
   \hline
    \end{tabular}
    \caption{
    Timescales estimated for the $\Lambda$-hyperon equilibration, $\tau_\Lambda$, at conditions~\eqref{eq:typicalPNS}. These estimates are obtained using the average rate in Eq.~\eqref{eq:tau_lambda_approx} and the approximations described in the main text. The table includes nonleptonic scattering channels and the two nonleptonic coalescence channels in Eq.~\eqref{eq:HADcoal_channels_explicit}.}
    \label{tab:HADtimescales}
\end{table}

\section{Semileptonic channels and neutrino transport}
\label{sec:SL_transport}

The nonleptonic rates discussed above determine the chemical equilibration of $\Lambda$ hyperons in the PNS. Semileptonic channels are much slower and therefore do not control the equilibrium abundance. Their relevance is instead different: once a thermal population of $\Lambda$ hyperons is present, these reactions open additional charged-current absorption channels for neutrinos and antineutrinos. In this section we therefore evaluate the semileptonic rates both as a consistency check on their subleading role in equilibration and, more importantly, as inputs for the corresponding neutrino opacities.

Semileptonic reactions are induced by the fundamental interactions 
$u +  \ell^- \leftrightarrows s + \nu_\ell$ and $u + \bar \nu_\ell \leftrightarrows s + \ell^+$. We distinguish between semileptonic scattering processes,
\begin{align} 
\label{eq:SLscat}
    \ell^- + B  &\leftrightarrows  \nu_\ell + B^\prime~, 
    \nonumber\\
    \bar\nu_\ell  + B &\leftrightarrows \ell^+ + B^\prime~,
\end{align}
and the semileptonic coalescence process,
\begin{align}
\label{eq:SLcoal}
    B + \ell^- + \bar\nu_\ell &\leftrightarrows B^\prime~.
\end{align}

\subsection{Semileptonic scattering}
\label{sec:SLscat_rates}

As a first semileptonic contribution to $\Lambda$ production, we consider the scattering channels listed in Eq.~\eqref{eq:SLscat}:
\begin{align}
p + \ell^- & \longrightarrow \Lambda + \nu_\ell~,
\nonumber\\
p + \bar\nu_\ell & \longrightarrow \Lambda + \ell^+~,\qquad \ell=e,\mu~.
\end{align}
To evaluate the corresponding rates, we start from the weak effective Lagrangian
\begin{equation}
\label{eq:weakLag}
    \mathcal L_{W}^\ell \supset -\frac{\GF V_{us}}{\sqrt{2}}\,\left(\bar{s}\gamma_\mu (1-\gamma_5)u\right)\,\left(\bar\nu_\ell \gamma^\mu (1-\gamma_5)\ell\right) +\text{h.c.},
\end{equation}
where $V_{us}$ is the CKM matrix element for the transition between the $s$ and $u$ quarks with $|V_{us}|=0.22431(85)$ ~\cite{ParticleDataGroup:2024cfk}. Neglecting electromagnetic $\mathcal O(\alpha)\approx1/137$ corrections, the amplitude of the processes in Eq.~\eqref{eq:SLscat} is factorized into a leptonic and a hadronic current, where the latter is parametrized by baryonic form factors~\cite{Cabibbo:2003cu,Weinberg:1958ut}
\begin{align}
 &&\langle B^\prime (p^\prime) | \bar{u} \gamma_\mu s | B (p)\rangle 
 =
 \bar{u}^\prime (p^\prime)  \Big[
 f_1(q^2)  \,  \gamma_\mu    
 \nonumber\\
 &&+ \frac{f_2(q^2)}{M}   \, \sigma_{\mu \nu}   q^\nu  
 + \frac{f_3(q^2)}{M}   \,  q_\mu  
 \Big]  
  u (p),  \label{eq:vectorFF}\\ 
 &&\langle B^\prime (p^\prime) | \bar{u} \gamma_\mu \gamma_5  s |B (p)\rangle 
 =
 \bar{u}^\prime (p^\prime)  \Big[
 g_1(q^2)    \gamma_\mu    \nonumber\\
 && +\frac{g_{2} (q^2)}{M}   \sigma_{\mu \nu}   q^\nu  
 + 
 \frac{g_{3} (q^2)}{M}   q_\mu  
 \Big]   \gamma_5  u (p)\,.  \label{eq:axialvectorFF}
\end{align}
In this equation, $u$ are baryon spinor amplitudes, $M^{\prime}$ is the mass of the baryon $B^{\prime}$ and the form factors $f_i$ and $g_i$ depend on $q^2$, where $q = p - p^\prime$ is the momentum transfer. To simplify the hadronic matrix element and the rates, we adopt a double expansion in the SU(3)-flavor breaking parameter $\delta=(M-M^\prime)/M$ and in $q^2/M^2$, truncating at leading order. This leaves only the vector and axial form factors, $f_1(0)$ and $g_1(0)$, neglecting $f_{2,3}$, $g_{2,3}$ and the residual $q^2$ dependence. For the hyperonic channels, we use the corresponding leading-order couplings that are obtained from the standard SU(3)-flavor relations, where the vector coupling is protected against first-order SU(3)-breaking corrections due to the Ademollo--Gatto theorem~\cite{Ademollo:1964sr}. For the non-strange $n \longrightarrow p$ transition, this yields $f_1^{np}(0)=1$, and $g_1^{np}(0)=g_A$, while for the strange $\Lambda \longrightarrow p$ one,  $f_1^{\Lambda p}(0)=-\sqrt{3/2}$ and $g_1^{\Lambda p}(0)=-1/\sqrt{6}(D+3F)$~\cite{Cabibbo:2003cu}, where $D$, $F$ and $g_A$ are baryon axial couplings introduced in Sec.~\ref{sec:onemeson}.

\begin{figure*}[t!]
    \centering
\begin{tabular}{cc}    \includegraphics[width=0.5\linewidth]{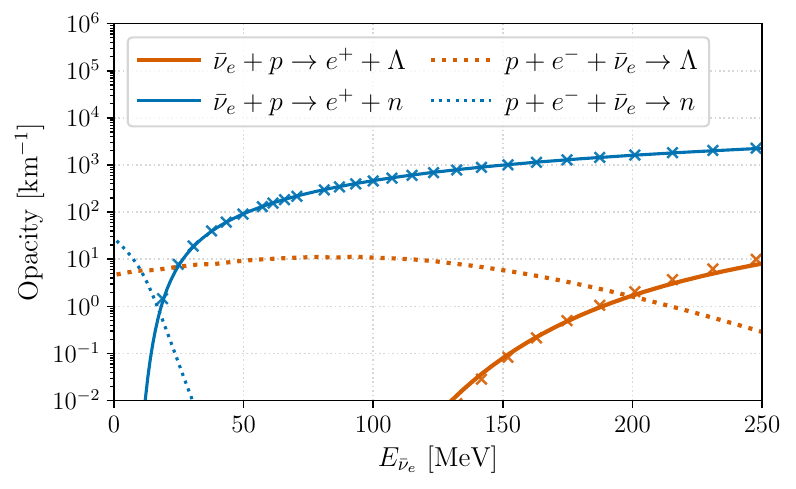} & \includegraphics[width=0.5\linewidth]{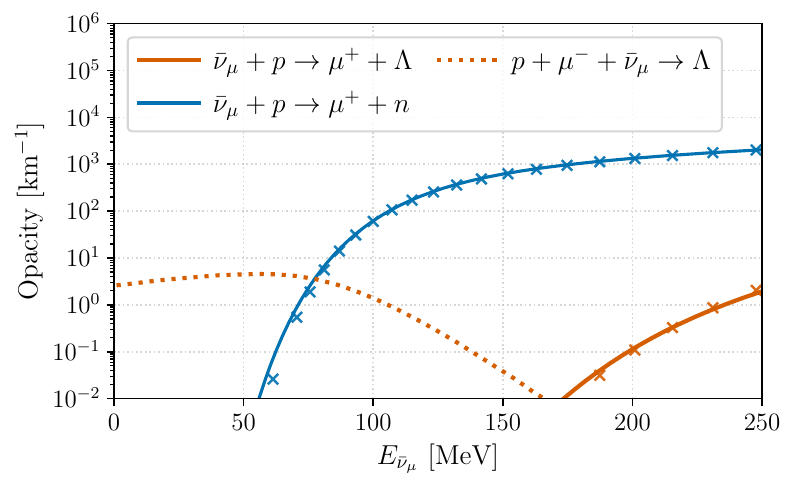}\\   \includegraphics[width=0.5\linewidth]{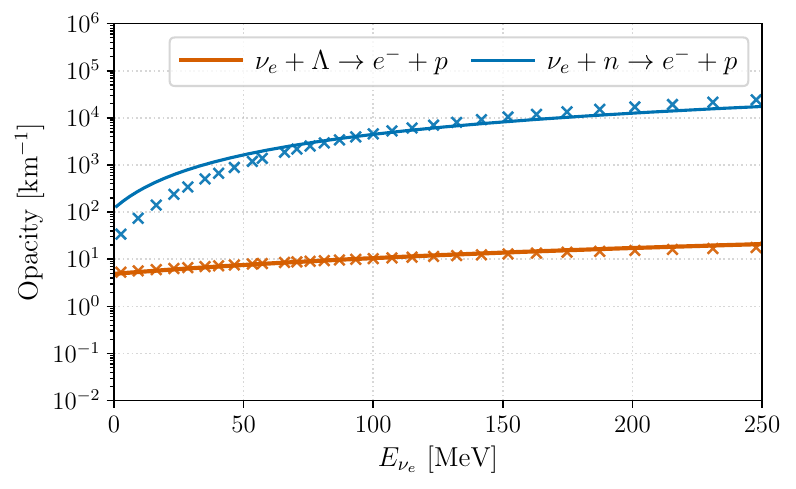} & \includegraphics[width=0.5\linewidth]{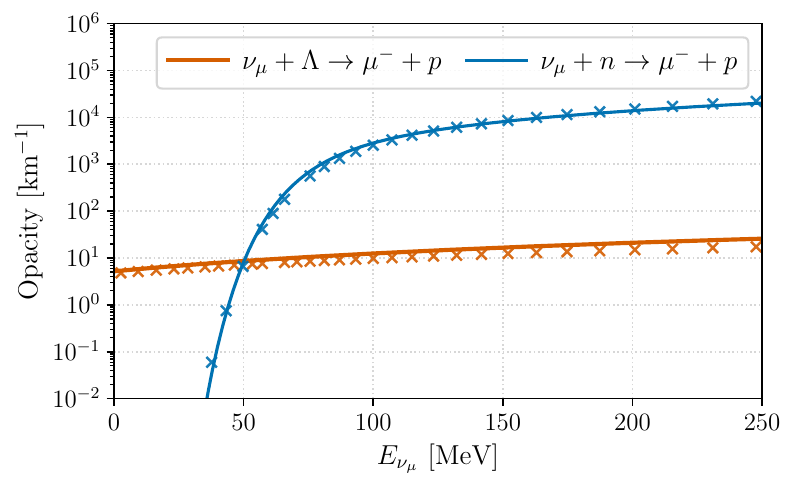}
\end{tabular}
\caption{(Anti)neutrino opacities induced by $\Lambda$-hyperons using the PNS in Eq.~\eqref{eq:typicalPNS}. Orange and blue lines denote the $\Lambda$ and neutron channels, respectively. For both colors, solid lines show semileptonic scattering contributions, while dotted lines show semileptonic coalescence contributions. Crosses indicate the corresponding opacities from the full calculation of Ref.~\cite{Guo:2020tgx}, discussed in the main text.}
\label{fig:opacities}
\end{figure*}

At leading order, the semileptonic amplitudes are evaluated at tree level from Eq.~\eqref{eq:weakLag}, retaining only a reduced subset of the form factors introduced above. The corresponding spin-summed squared matrix elements are then used to construct the cross sections entering Eq.~\eqref{eq:RateScat} for semileptonic scattering. We use the same approximations for the Pauli blocking factors and baryon dispersion relations as in the calculation of nonleptonic scattering rates. In this case, a direct comparison is possible with the benchmark opacity treatment of Ref.~\cite{Guo:2020tgx}, which uses a fully inelastic phase-space calculation and includes higher-order contributions to the hadronic current, such as weak magnetism, pseudoscalar form factors, and mean-field medium modifications (see also Refs.~\cite{Shen2012PhRvC86,Roberts:2017}).\footnote{An extension to the mean field approximation for the dressing of the baryons in these calculations has been recently provided in the relativistic Hartree-Fock approach, featuring explicit momentum dependent nucleon self energies~\cite{Sokolowski2026arXiv260517563S} (see also Ref.~\cite{Reddy99} for the role of many-body correlations, which modify the opacity substantially in excess of nuclear saturation density.)} To perform the comparison, we compute the corresponding electronic and muonic weak processes, including the inverse neutron decay, based on the numerical 2D-integrals approach of Ref.~\cite{Guo:2020tgx}, with the proper replacements of $V_{ud}\longrightarrow V_{us}$ and of vector and axial-vector coupling constants shown above for the case of $\Lambda \longrightarrow p$ transition. As shown in Fig.~\ref{fig:opacities}, our approximate results (solid lines) agree quantitatively with that calculation of Ref.~\cite{Guo:2020tgx} implementing the full kinematics (marked by $\times$ symbols), except for the process $n + \nu_e \longrightarrow p + e^-$, where our approximate treatment underestimates the opacity at high neutrino energy and overestimates the opacity at $E_{\nu_e}<70$ MeV.

\subsection{Semileptonic coalescence} 
\label{sec:SLcoal}
As a second semileptonic contribution to $\Lambda$ production, we consider the semileptonic coalescence channels listed in Eq.~\eqref{eq:SLcoal}:
\begin{align}
p + \ell^- + \bar\nu_\ell & \longrightarrow \Lambda~,\qquad \ell=e,\mu~.
\label{eq:SLcoal_channels}
\end{align}
For these processes, we adopt the formulation of Ref.~\cite{Guo:2020tgx}, in which the antineutrino absorption rate is evaluated without invoking the Pauli-blocking and baryon-dispersion approximations in Eqs.~\eqref{eq:Fdeg} and \eqref{eq:ApproxDispRel}. It can be written as a two-dimensional integral over energies:
\begin{equation}
    \begin{aligned}
    \label{eq:RateSLcoal}
    \Gamma_{\bar \nu}(E_{\bar \nu})
    &= \frac{G_F^2 V_{us}^2}{4\pi^3 E_{\bar \nu}^2}
        \int {\rm d}E_B \int {\rm d}E_\ell\;
        f_B\, f_\ell\, (1-f_{B^\prime})\,\\
        &
        \qquad\qquad\times\,\Bigl[
        \mathcal A\, I_{\mathcal A}
        + \mathcal B\, I_{\mathcal B}
        + \mathcal K\, I_{\mathcal K}
        \Bigr]\,.
    \end{aligned}
\end{equation}
The quantities $I_i$, with $i\in\{\mathcal A, \mathcal B, \mathcal K\}$, are angular integrals (discussed in App.~\ref{app:main}) obtained by analytically integrating out the phase-space angles in the squared amplitude, while the coefficients multiplying them encode the dependence on the weak form factors $(f_1,\,g_1)$ introduced above:
\begin{equation}
    \begin{aligned}
        \mathcal A &= (g_1 + f_1)^2,\quad
        \mathcal B = (g_1 - f_1)^2, \quad\\[3pt]
        \mathcal K &= (g_1^2 - f_1^2) M M^\prime\,.
    \end{aligned}
\end{equation}
The corresponding collision operator is obtained by integrating the rate over the antineutrino phase space, as in Eq.~\eqref{eq:collOp}. 

\begin{table}
\renewcommand{\arraystretch}{1.7}
\setlength{\arrayrulewidth}{.25mm}
  \setlength{\tabcolsep}{0.5 em}
    \centering
    \begin{tabular}{|c|c|}
    \hline
    Reaction     & $\tau_\Lambda$ ($10^{-6}$ s) \\
    \hline
    $p + \bar\nu_e \longrightarrow \Lambda + e^+$  & 2.0 \\
    $p + \bar\nu_\mu \longrightarrow \Lambda + \mu^+$ & 2.8\\
    \hline
   $p + e^- \longrightarrow \Lambda + \nu_e$ & 0.26 \\
   $p + \mu^- \longrightarrow \Lambda+ \nu_\mu$ & 0.34 \\
   \hline
   $p + e^- + \bar{\nu}_e \longrightarrow \Lambda$ & 0.37\\
   $p+ \mu^- + \bar{\nu}_\mu \longrightarrow \Lambda$& 1.15\\
   \hline
    \end{tabular}
    \caption{Timescales estimated for the $\Lambda$-hyperon equilibration, $\tau_\Lambda$, at the conditions~\eqref{eq:typicalPNS}. These estimates are obtained using the average rate in Eq.~\eqref{eq:tau_lambda_approx} and the approximations described in the main text.}
    \label{tab:timescales}
\end{table}

In Tab.~\ref{tab:timescales} we report the estimated equilibration timescales of $\Lambda$ hyperons from semileptonic processes under PNS conditions~\eqref{eq:typicalPNS}. 
The dominant channels are the semileptonic scattering processes 
$p + e^- \longrightarrow \Lambda + \nu_e$ 
and 
$p + \mu^- \longrightarrow \Lambda + \nu_\mu$, 
with the electronic channel providing the largest contribution because of the high electron abundance. For the same reason, 
$p + e^- + \bar{\nu}_e \longrightarrow \Lambda$ 
also gives a significant contribution, even though it is a three-body coalescence process.
The semileptonic equilibration timescales are nevertheless about four orders of magnitude longer than those associated with nonleptonic reactions, which drive $\Lambda$ hyperons to chemical equilibrium almost instantaneously on PNS dynamical timescales. Their contribution to establishing local $\Lambda$ chemical equilibrium is therefore negligible compared with the nonleptonic one.

It is useful to recall that a similar hierarchy between nonleptonic and semileptonic processes already exists in vacuum $\Lambda$ decays. The two-body nonleptonic channels  $\Lambda \longrightarrow N + \pi$  nearly saturate the decay width, while the semileptonic modes are suppressed by roughly three orders of magnitude, due to their different phase-space and $Q$-value scaling. The corresponding coalescence processes seem to retain this pattern also in a hot and dense medium, as can be seen by comparing Tabs.~\ref{tab:HADtimescales} and \ref{tab:timescales}.

The more relevant implication of these channels is instead for neutrino transport. In Fig.~\ref{fig:opacities} we show the (anti)neutrino opacities from semileptonic scattering and coalescence for conditions~\eqref{eq:typicalPNS}, with the $\Lambda$-induced contributions highlighted in orange. The top panels show the antineutrino opacities, including both semileptonic scattering and coalescence channels, separately for electron and muon flavors. The bottom panels show the contribution to neutrino opacities from the corresponding scattering processes. 

Overall, the (anti)neutrino opacities induced by $\Lambda$ hyperons are approximately three orders of magnitude smaller than those from standard $\beta$ reactions involving neutrons. This suppression arises from two factors: (1) the much lower abundance of hyperons relative to neutrons (relevant for neutrino absorption), and (2) the additional reduction of hyperonic rates by the factor $|V_{us}|^2 \simeq 0.04$.

An exception is found for low-energy muonic neutrinos and antineutrinos. For $E_{\nu_\mu} \lesssim 50$ MeV, muon production requires high-energy neutrons, whose population lies in the Boltzmann-suppressed tail of the distribution. However, because the mass difference between $\Lambda$ and the proton is larger than the muon mass, neutrinos can be absorbed by hyperons even at threshold, opening a new absorption channel for low-energy muonic neutrinos.

A similar effect appears for antineutrino opacity through the coalescence process. The opening of the inverse $\Lambda$-hyperon decay channel is particularly relevant for the absorption of low-energy muonic antineutrinos in the range $E_{\bar \nu_\mu} \lesssim 80$~MeV. This channel has no direct neutron counterpart, since the corresponding muonic coalescence process $p + \mu^- + \bar\nu_\mu \longrightarrow n$ is forbidden by energy-momentum conservation.

By contrast, the presence of $\Lambda$ hyperons has little impact on neutrino and antineutrino opacities in the electronic sector. This is because low-energy electronic neutrino absorption on neutrons is not impeded by any kinematic threshold. Likewise, electronic antineutrino absorption through coalescence on a $\Lambda$ hyperon is less significant, since antineutrinos can be more efficiently absorbed by protons and through inverse neutron decay.

The appearance of these new low-energy absorption channels for neutrinos and antineutrinos suggests that the most relevant implication of $\Lambda$-induced semileptonic reactions may be on neutrino transport. Specifically, our results identify two additional low-energy muonic absorption channels, one for $\nu_\mu$ and one for $\bar\nu_\mu$, which could induce nontrivial effects on the muonization of PNS matter (see Ref.~\cite{Fischer:2020vie}), competing with purely leptonic processes involving muons, e.g., the inverse muon decay channel, $e^- + \bar\nu_e + \nu_\mu \longrightarrow \mu^-$ and other channels such as $\nu_e + e^- \longrightarrow \mu^- + \nu_\mu$ and $\nu_\mu + e^- \longrightarrow \mu^- + \nu_e$, all of which have low-energy opacities similar to those of $\nu_\mu +\Lambda\longrightarrow \mu^- + p$ and $p+\mu^-+\bar\nu_\mu\longrightarrow\Lambda$, on the order of $\chi\simeq 1$--$10$~km$^{-1}$ (see Fig.~\ref{fig:opacities}), as well as corresponding antineutrino reactions for the production of $\mu^+$ such as $\bar\nu_\mu + e^+ \longrightarrow \mu^+ + \bar\nu_e$ (see Figs.~5 and 6 in Ref.~\cite{Guo:2020tgx}). 

A similar result was found in Ref.~\cite{Fore:2019wib}, where thermal $\pi^-$ in hot and dense matter were shown to open additional low-energy $\nu_\mu$ and $\bar\nu_\mu$ absorption channels through the charged-current processes $\pi^- + \nu_\mu \longrightarrow \mu^-$ and $\mu^- + \bar\nu_\mu \longrightarrow \pi^-$. However, the impact of these processes strongly depends on the in-medium dispersion relation of pions (which is not yet fully understood)~\cite{Fore2024PhRvC110}, since the process $\pi^- + \nu_\mu \longrightarrow \mu^-$ is not kinematically allowed in vacuum. 
While the $\Lambda$-induced $\nu_\mu$ and $\bar{\nu}_\mu$ absorption channels we presented do not suffer from that uncertainty, a quantitative assessment of this effect requires fully-fledged CCSN simulations featuring Boltzmann neutrino transport including the $\Lambda$-induced opacities presented above.

\section{Conclusions} \label{sec:Conclusions}

In this work, we demonstrated for the first time that the emergence of $\Lambda$ hyperons under typical post-bounce PNS conditions is controlled by nonleptonic weak interaction channels. In particular, the scattering channels, $N+N\longrightarrow N + \Lambda$, 
drive the fastest equilibration, with characteristic timescales of order $(0.01$--$0.1)\times 10^{-9}$~s, while the coalescence channels,
$\pi + N\leftrightarrows\Lambda$, are slower but still on the order of $\lesssim 10^{-9}$~s. These timescales are much shorter than any characteristic evolution timescales of the PNS, supporting the use of locally chemically equilibrated $\Lambda$ abundances under the thermodynamic conditions considered here.

A related methodological point is that short-distance contributions to the $N+N\leftrightarrow N+\Lambda$ transition can be sizable using the EFT framework of Refs.~\cite{Parreno:2003ny,Parreno:2003nx}, where one-meson-exchange interactions are supplemented by four-baryon contact operators fitted to hypernuclear nonmesonic weak-decay data. In our implementation, the contact terms of Eq.~\eqref{eq:CTs}, with the LECs in Eq.~\eqref{eq:LECsdata}, dominate over the long-range OPE and OKE contributions under the PNS conditions studied here. This indicates that short-range weak physics plays an important role in hot dense matter, with implications for microphysical inputs to hyperonic bulk-viscosity studies in neutron-star matter and merger remnants~\cite{Alford:2020pld,vanDalen:2003uy}.

By contrast, semileptonic channels are subdominant for setting $\Lambda$ chemical equilibration, although they remain relevant for neutrino transport. In particular, $\Lambda$-induced processes open additional low-energy absorption channels in the muonic sector. For the assumed conditions in the PNS core, in the energy windows $E_{\nu_\mu}\lesssim50$ MeV and $E_{\bar\nu_\mu}\lesssim80$ MeV, the corresponding $\Lambda$-induced neutrino opacity exceeds that due to matter containing only neutrons and protons. By contrast, the analogous effect in the electronic sector results comparatively small.

Overall, our results point to a potentially relevant role of hyperonic microphysics for flavor-dependent neutrino opacities and, consequently, for PNS muonization. A quantitative assessment of this impact requires fully self-consistent CCSN simulations including hyperonic microphysics and neutrino transport. A systematic assessment of the dependence on the EOS is also an important goal for future studies, since both the equilibrium hyperon abundance and the associated weak rates can be sensitive to the dense-matter composition.

\section*{Acknowledgments}
RZ, JMC and PDS acknowledge support from the European Union through the grant ``UNDARK'' of the Widening participation and spreading excellence programme (project number 101159929). RZ and JMC also acknowledge the MICINN through the grant ``DarkMaps'' PID2022-142142NB-I00. 
TF was supported by the Polish National Science Center (NCN) under Grant No. 2023/49/B/ST9/03941. The computations for the neutrino opacity in the full kinematics approach were performed at the Wroclaw Center for Scientific Computing and Networking (WCSS). This work made use of {\tt FeynCalc}. We acknowledge the developers of {\tt FeynCalc} for providing this tool~\cite{Shtabovenko_2025,MERTIG1991345}. 
     
\onecolumngrid
\appendix

\section{Inputs for the one-meson-exchange contributions}
\label{app:AB_extraction}

In this appendix we summarize the extraction of the weak couplings $\mathcal A$ and $\mathcal B$ entering the effective interaction in Eq.~\eqref{eq:LagWeak}. These are obtained from data on nonleptonic hyperon decays, where the pion is the final state. The matrix element is
\begin{equation}
\mathcal M_{\phi B^\prime B} =i\,
G_F m_{\pi^+}^2\,\bar u^\prime\,(\mathcal A_{\phi B^\prime B}-\mathcal B_{\phi B^\prime B}\gamma_5)\,u,
\label{eq:app_M_AB}
\end{equation}
where $\phi$ is a pion and we use the experimental $B \longrightarrow B^\prime + \phi$ rates and decay parameters from the PDG~\cite{ParticleDataGroup:2024cfk} to extract the corresponding values of the amplitudes.

\subsection{Partial widths and asymmetry parameters}
In the $B$ rest frame, the final baryon momentum and energy are respectively
\begin{align}
|\vec p| &= \frac{\lambda_{\rm K}^{1/2}(M^2,M^{\prime\,2},m_{\phi}^2)}{2M}, \qquad
E= \frac{M^{\prime\,2}+M^2-m_{\phi}^2}{2M}\,.
\label{eq:app_kinAB}
\end{align}
Using Eq.~\eqref{eq:app_M_AB}, the corresponding two-body decay width is
\begin{equation}
\Gamma=
\frac{G_F^2 m_{\pi^+}^4}{4\pi M}\,
|\vec p|(E+M^\prime)
\left[
|\mathcal A_{\phi B^\prime B}|^2+\frac{E-M^\prime}{E+M^\prime}|\mathcal B_{\phi B^\prime B}|^2
\right]\,,
\label{eq:app_width_AB}
\end{equation}
which has to be matched with its experimental value,
\begin{equation} \label{eq:app_GammaBR}
    \Gamma = \frac{{\rm BR}}{\tau}\,,
\end{equation}
where $\mathrm{BR}$ denotes the branching ratio. This procedure fixes the overall normalization of the weak hadronic couplings.
The quantities $\mathcal A_{\phi B^\prime B}$ and $\mathcal B_{\phi B^\prime B}$ are assumed to be real, since complex phases, either strong phases from rescattering effects or weak phases from CP violation, are small.

A second independent observable is provided by the decay asymmetry parameter \(\alpha\), which probes the interference between the \(S\)- and \(P\)-wave contributions. In the standard notation one has
\begin{equation}
s=\mathcal A_{\phi B^\prime B},
\qquad
p=\frac{|\vec p|}{E+M^\prime}\,\mathcal B_{\phi B^\prime B}.
\label{eq:app_spdef}
\end{equation}
With this decomposition, the decay asymmetry parameter is usually defined as
\begin{equation}
\alpha=\frac{2\,\mathrm{Re}(s^*p)}{|s|^2+|p|^2}.
\label{eq:app_alpha_def}
\end{equation}
which, assuming $\mathcal A_{\phi B^\prime B}$ and $\mathcal B_{\phi B^\prime B}$ real, becomes
\begin{equation}
\alpha=\frac{2r}{1+r^2},
\qquad
r\equiv \frac{|\vec p|}{E+M^\prime}\frac{\mathcal B_{\phi B^\prime B}}{\mathcal A_{\phi B^\prime B}}.
\label{eq:app_alpha_r}
\end{equation}
which gives an Equation~\eqref{eq:app_alpha_r} with two branches,
\begin{equation}
r^{(\pm)}=\frac{1\pm\sqrt{1-\alpha^2}}{\alpha}.
\label{eq:app_r_branches}
\end{equation}
We select the $S$-wave-dominated branch, i.e. the solution with smaller $|r|$.

\subsection{Extraction of $\mathcal A$ and $\mathcal B$}
Defining
\begin{equation}
K=
\frac{G_F^2 m_{\pi^+}^4}{4\pi M}\,
|\vec p|(E+M^\prime),
\label{eq:app_Kdef}
\end{equation}
the corresponding weak hadronic couplings for the decay $B \longrightarrow B^\prime + \phi$ are then given by
\begin{equation}
\mathcal A_{\phi B^\prime B}=\pm \sqrt{\frac{\Gamma/K}{1+r^2}},
\qquad
\mathcal B_{\phi B^\prime B}=\frac{E+M^\prime}{|\vec p|}\,r\,\mathcal A_{\phi B^\prime B},
\label{eq:app_A_B_extraction}
\end{equation}
where $\Gamma$ is taken from Eq.~\eqref{eq:app_GammaBR} and $r$ is obtained from Eq.~\eqref{eq:app_r_branches} using the measured value of $\alpha$. Since the decay width and asymmetry parameter leave an overall sign undetermined, an additional convention is needed to fix the signs. We choose the PDG criterion which is largely based on measurements of a third decay parameter $\gamma=|s|^2-|p|^2/|s|^2+|p|^2$, which breaks the degeneracy~\cite{Lee:1957qs,ParticleDataGroup:2024cfk}.

\begin{table}[htbp]
\renewcommand{\arraystretch}{1.7}
\setlength{\arrayrulewidth}{.25mm}
  \setlength{\tabcolsep}{0.5 em}
\centering
\begin{tabular}{|lccc|}
\hline
Decay              & $\mathcal{A}_{\phi B^\prime B}$ & $\mathcal{B}_{\phi B^\prime B}$ & $|\mathcal{B}_{\phi B^\prime B}/\mathcal{A}_{\phi B^\prime B}|$ \\
\hline
$\Lambda \to p + \pi^-$   & $1.387(8)$   & $11.7(2)$    & 8.4(1) \\
$\Lambda \to n + \pi^0$   & $-1.044(8)$  & $-7.3(2)$     & 6.9(2) \\
$\Sigma^+ \to p + \pi^0$  & $1.44(3)$    & $-11.6(3)$    & 8.3(7) \\
$\Sigma^+ \to n + \pi^+$  & $0.044(2)$   & $18.56(7)$   & $419(2)$ \\
$\Sigma^- \to n + \pi^-$  & $1.879(7)$   & $-0.63(7)$    & 0.33(4) \\
$\Xi^0 \to \Lambda + \pi^0$ & $1.52(2)$  & $-4.8(2)$     & 2.98(8) \\
$\Xi^- \to \Lambda + \pi^-$ & $1.993(9)$ & $-6.4(1)$     & 3.11(5) \\
\hline
\end{tabular}
\caption{Extracted $\mathcal{A}_{\phi B^\prime B}$ (PV) and $\mathcal{B}_{\phi B^\prime B}$ (PC) 
amplitudes for spin-$1/2$ hyperon nonleptonic decays, using PDG~\cite{ParticleDataGroup:2024cfk} data and assuming real amplitudes. Amplitudes in units of $G_F m_{\pi^+}^2$.}
\label{tab:hyp_amplitudes}
\end{table}

One specific ingredient in our calculation, related to the weak kaon-nucleon vertex appearing in the OKE contributions, cannot be extracted directly from data. To predict it, we use an SU(3)-flavor chiral Lagrangian that relates the different contributions to $\mathcal A_{\phi B^\prime B}$ and $\mathcal B_{\phi B^\prime B}$. We do not present the full formalism here, but instead follow the notation and prescriptions introduced in Ref.~\cite{Jenkins:1991bt}, where the leading-order weak chiral Lagrangian is parametrized in terms of the two LECs $h_D$ and $h_F$, written in units of $G_F m_{\pi^+}^2f_\pi$. In Tab.~\ref{tab:hyp_amplitudes} we show the updated values of $\mathcal A_{\phi B^\prime B}$ and $\mathcal B_{\phi B^\prime B}$ obtained using current data on nonleptonic hyperon decays from the PDG~\cite{ParticleDataGroup:2024cfk}. A fit to these data approximately yields $h_D\simeq -0.80$ and $h_F\simeq1.92$, which are in the ballpark of the results of the fit in Ref.~\cite{Jenkins:1991bt} to the older data. 

The parity-violating amplitudes at leading order  are simple linear combinations of the LECs,
\begin{align}
\label{eq:kaon_pv_amplitudes}
\mathcal A_{K^0pp}=\frac{h_F-h_D}{\sqrt{2}}=1.93,\;\;\mathcal A_{K^+np}=\frac{h_D+h_F}{\sqrt{2}}=0.80,
\end{align}
and $\mathcal A_{K^0nn}=\mathcal A_{K^0pp}+A_{K^+np}=2.73$.
The parity-conserving amplitudes are obtained from baryon-pole contributions induced by a combination of weak and strong vertices~\cite{Bijnens:1985kj,Jenkins:1991bt}. For the weak kaon-nucleon amplitudes one obtains
\begin{align}
\label{eq:baryonpole}
&\mathcal B_{K^0pp}=\sqrt{2}(h_F-h_D)(D-F)\frac{M_N}{M_\Sigma-M_N}=4.83,\nonumber\\
&\mathcal B_{K^+np}=\frac{(h_D-h_F)(D-F)}{\sqrt{2}}\frac{M_N}{M_\Sigma-M_N}-\frac{(h_D+3h_F)(D+3F)}{3\sqrt{2}}\frac{M_N}{M_\Lambda-M_N}=-16.00,
\end{align}
with $\mathcal B_{K^0nn}=\mathcal B_{K^0pp}+\mathcal B_{K^+np}=-11.17$ and where we have used isospin averages for the baryon masses. It is important to emphasize that, once the LECs are fixed using the parity-violating amplitudes, the leading-order chiral predictions for the parity-conserving amplitudes fail to reproduce the data of nonleptonic hyperon decays satisfactorily~\cite{Bijnens:1985kj,Jenkins:1991bt}. Consequently, the values reported in Eq.~\eqref{eq:baryonpole} should be regarded as subject to large uncertainties.

\section{Matrix elements used in the rate calculations}
\label{app:SL_sqme}

In this appendix we list the matrix elements corresponding to the processes listed in the main text in Eqs.~\eqref{eq:HADscat},\eqref{eq:HADcoal},\eqref{eq:SLscat} and \eqref{eq:SLcoal}. 

\subsection{Semileptonic processes}

For the semileptonic scattering channel
\begin{align}
\ell^- + B &\longrightarrow \nu_\ell + B'
\label{eq:app_SL_channels_scatt}
\end{align}
where $B$ and $B'$ denote octet baryons connected by a strangeness-changing $\Delta S=-1$ weak charged current. With the approximation at leading order in SU(3)-flavor breaking and $q^2/M^2$ explained in the main text,
the spin-averaged squared matrix element reads
\begin{align}
\overline{|\mathcal M|^2}_{\ell^- B\to \nu_\ell B'}
= 8\,G_F^2\,|V_{us}|^2\,
\Big[
(f_1+g_1)^2\,(p_{\nu_\ell}\!\cdot p_B)(p_{\ell^-}\!\cdot p_{B'})&
+(f_1-g_1)^2\,(p_{\nu_\ell}\!\cdot p_{B'})(p_{\ell^-}\!\cdot p_B)\nonumber\\
&\qquad
+(g_1^2-f_1^2)\,M_BM_{B'}\,(p_{\nu_\ell}\!\cdot p_{\ell^-})
\Big]\,.
\label{eq:app_SL_M2_nu}
\end{align}

All other semileptonic channels used in this work are obtained from this expression, by time reversal and/or crossing of the external lepton lines. For example, the time-reversed process of
\(\ell^- + B \longrightarrow \nu_\ell + B'\) is
\begin{equation}
\nu_\ell + B' \longrightarrow \ell^- + B .
\end{equation}
The spin-summed matrix element is the same as for Eq.~\eqref{eq:app_SL_channels_scatt}, with initial and final momenta interchanged. The corresponding spin-averaged quantity is larger by a factor of two, because the incoming neutrino carries only
one helicity state, whereas the incoming charged lepton in Eq.~\eqref{eq:app_SL_M2_nu} was averaged over two spin states.

The spin-summed matrix element of the other semileptonic scattering channel
\begin{equation}
\bar\nu_\ell + B \longrightarrow \ell^+ + B',
\end{equation}
is obtained by crossing the charged-lepton and neutrino lines of Eq.~\eqref{eq:app_SL_channels_scatt}.
Finally, the semileptonic coalescence channel
\begin{equation}
B + \ell^- + \bar\nu_\ell \longrightarrow B'
\end{equation}
is obtained by crossing only the final neutrino line.

\subsection{Nonleptonic processes}

Let us now consider the one-meson-exchange (OME) contributions. As discussed in the main text, both OPE and OKE amplitudes are built from one strong and one weak baryon--baryon--meson vertex. For the nonleptonic scattering channels $n + n \longrightarrow n + \Lambda$ and $p + n \longrightarrow p + \Lambda$, each exchanged meson gives rise to two topologies (see Figs.~\ref{fig:OPE} and \ref{fig:OKE}). Thus, each channel receives four OME contributions, which must be summed at the amplitude level. Correctly tracking the relative signs and coefficients of these terms is therefore essential, since they determine the
interference pattern in the spin-averaged squared matrix element
\(\overline{|\mathcal M|^2}\).

As a concrete example, let us spell out the construction for the channel
\begin{equation}
    \label{eq:app_pn_to_pL}
    p(p_1)+n(p_2)\to p(p_3)+\Lambda(p_4).
\end{equation}
The Mandelstam variables are defined with this fixed external-state ordering, so that the direct topology carries momentum transfer ($q^2=t$), whereas the charge-exchange topology carries ($q^2=u$).
For pion exchange we write
\begin{equation}
\mathcal M^{\rm ope}_{pn}
=
\mathcal M^{\rm ope}_{pn,t}
+
\mathcal M^{\rm ope}_{pn,u} ,
\end{equation}
with
\begin{align}
\mathcal{M}^{\rm ope}_{pn,t}
&=
-G_F m_{\pi^+}^2\,\frac{(D+F)M_p}{f_\pi}\,
\frac{
\left[\bar u_p(p_3)\gamma_5 u_p(p_1)\right]
\left[\bar u_\Lambda(p_4)
\left(\mathcal A_{\pi^0 n \Lambda}+\mathcal B_{\pi^0 n \Lambda}\gamma_5\right)
u_n(p_2)\right]
}{
t-m_{\pi^0}^2
},
\\[1ex]
\mathcal{M}^{\rm ope}_{pn,u}
&=
+\sqrt{2}\,G_F m_{\pi^+}^2\,\frac{(D+F)(M_n + M_p)}{2f_\pi}\,
\frac{
\left[\bar u_p(p_3)\gamma_5 u_n(p_2)\right]
\left[\bar u_\Lambda(p_4)
\left(\mathcal A_{\pi^- p \Lambda}+\mathcal B_{\pi^- p \Lambda}\gamma_5\right)
u_p(p_1)\right]
}{
u-m_{\pi^+}^2
},
\end{align}
where $M_n$ and $M_p$ are the neutron and proton masses, respectively, and we have used that $\mathcal C_{\pi^0 pp}=-\mathcal C_{\pi^0 nn}=\mathcal C_{\pi^-pn}/\sqrt{2}=D+F$, in the notation of the strong vertex in Eq.~\eqref{eq:LagStrong}. The numerical values of the weak PV and PC amplitudes entering these diagrams are listed in Tab.~\ref{tab:hyp_amplitudes}.

The first term corresponds to neutral-pion exchange: the strong vertex connects
$p(p_1)$ to $p(p_3)$, while the weak vertex connects $n(p_2)$ to
$\Lambda(p_4)$. The second term corresponds to charged-pion exchange: the
strong vertex connects $n(p_2)$ to $p(p_3)$, while the weak vertex connects
$p(p_1)$ to $\Lambda(p_4)$. The relative sign comes from vertex signs together with Wick-contraction, while the factor $\sqrt2$ comes from the strong vertex.

For kaon exchange the same channel receives
\begin{equation}
\mathcal M^{\rm oke}_{pn}
=
\mathcal M^{\rm oke}_{pn,t}
+
\mathcal M^{\rm oke}_{pn,u} ,
\end{equation}
with
\begin{align}
\mathcal{M}^{\rm oke}_{pn,t}
&=
+G_F m_{\pi^+}^2\frac{(D+3F)(M_\Lambda+M_n)}
{2\sqrt{3}f_\pi}\,
\frac{
\left[\bar u_p(p_3)
\left(\mathcal A_{K^0 p p}-\mathcal B_{K^0 p p}\gamma_5\right)
u_p(p_1)\right]
\left[\bar u_\Lambda(p_4)\gamma_5 u_n(p_2)\right]
}{
t-m_K^2
},
\\[1ex]
\mathcal{M}^{\rm oke}_{pn,u}
&=
-G_F m_{\pi^+}^2\frac{(D+3F)(M_\Lambda+M_p)}
{2\sqrt{3}f_\pi}\,
\frac{
\left[\bar u_p(p_3)
\left(\mathcal A_{K^- p n}-\mathcal B_{K^- p n}\gamma_5\right)
u_n(p_2)\right]
\left[\bar u_\Lambda(p_4)\gamma_5 u_p(p_1)\right]
}{
u-m_K^2
},
\end{align}
where we have used $\mathcal C_{K^0\Lambda n}=\mathcal C_{K^+\Lambda p}=-(D+3F)/\sqrt{3}$ for the strong vertex in Eq.~\eqref{eq:LagStrong}, while the PV and PC amplitudes entering the weak vertices are given in
Eqs.~\eqref{eq:kaon_pv_amplitudes} and \eqref{eq:baryonpole}.

The total OME amplitude is then
\begin{equation}
\mathcal M^{\rm ome}_{pn}
=
\mathcal M^{\rm ope}_{pn}
+
\mathcal M^{\rm oke}_{pn}.
\end{equation}

The channel $n + n \longrightarrow n + \Lambda$ also has two topology contributions $\mathcal M^{\rm ome}_{nn,t}$ and $\mathcal M^{\rm ome}_{nn,u}$ due to the two identical incoming neutrons, and it is treated analogously. In that case, however, the relative sign between the two topologies is fixed by Wick's theorem.

Finally, the nonleptonic coalescence processes considered here are the inverse of the
nonleptonic two-body decays
\begin{equation}
B \longrightarrow B' + \phi,
\end{equation}
with $\phi$ a pion. We therefore use the same effective weak couplings
$\mathcal A_{\phi B'B}$ and $\mathcal B_{\phi B'B}$ introduced in
Eq.~\eqref{eq:LagWeak}. For the inverse reaction
\begin{equation}
B' + \phi \longrightarrow B,
\label{eq:app_HAD_coal_generic}
\end{equation}
the production amplitude can be written, up to an overall phase, as
\begin{equation}
\mathcal M_{B'\phi\to B}
= -iG_Fm_{\pi^+}^2\,
\bar u_B
\left(
\mathcal A_{\phi B'B}
+
\mathcal B_{\phi B'B}\gamma_5
\right)
u_{B'}.
\label{eq:app_M_coal}
\end{equation}

Averaging over the spin of the incoming baryon $B'$ and summing over the spin
of the outgoing baryon $B$, one obtains

\begin{align}
\overline{|\mathcal M|^2}_{B'\phi\to B}
&=
2\,G_F^2\,m_{\pi^+}^4
\Big[
\left(
|\mathcal A_{\phi B'B}|^2
+
|\mathcal B_{\phi B'B}|^2
\right)
(p_{B'}\!\cdot p_B)+
\left(
|\mathcal A_{\phi B'B}|^2
-
|\mathcal B_{\phi B'B}|^2
\right)
M_{B'} M_B
\Big].
\label{eq:app_HAD_coal_generic_sq}
\end{align}

In the main text we specialize this result to the $\Lambda$-production channels
$p + \pi^- \longrightarrow \Lambda$ and $n+ \pi^0 \longrightarrow \Lambda$.

\subsection{Contact terms}
\label{app:CTs}
We start with the nonrelativistic $\Lambda + N \longrightarrow N + N$ potential at LO presented in~\cite{Parreno:2003ny,Parreno:2003nx},

\begin{align}
&V^{\rm LO}
  = \GF\,  \left(C_0^0+C_0^1\,\vec\sigma_1\cdot\vec\sigma_2\right)\left(C_{IS}+C_{IV}\,(\vec\tau_1\,h)\cdot\vec \tau_2\right),& h=\begin{pmatrix}0\\1\end{pmatrix},
\label{eq:VLO}
\end{align}
where we have labeled the line in which
$\Lambda \longleftrightarrow N$ occurs as 1 (with $h$ an isospin spurion), and as 2 the \textit{spectator line}. This can be matched to a nonrelativistic Hamiltonian,
\begin{align}
\label{eq:NRHamiltonian}
 \mathcal H_{\rm ct}
  = G_F\,\mathcal I_{ab;d}
  \left[
    C_0^0\,(N_a^\dagger\Lambda)(N_b^\dagger N_d)
    + C_0^1\,(N_a^\dagger\vec\sigma\Lambda)\cdot(N_b^\dagger\vec\sigma N_d)
  \right]+{\rm h.c.},
\end{align}
expressed in terms of two-component nucleon fields (see e.g. Ref.~\cite{Weinberg:1990rz,Weinberg:1991um,Hammer:2019poc}), and introduce the isospin tensor,
\begin{align}
\label{eq:isospin_tensor}
  \mathcal I_{ab;d}
  = C_{IS}\,h_a\delta_{bd}
    + C_{IV}\,(\tau^A h)_a\tau^A_{bd}.
\end{align}

Using the hermitian conjugate of Eq.~\eqref{eq:NRHamiltonian}, the charged-basis LO
Hamiltonian for $n + n \longrightarrow n + \Lambda$ is
\begin{align}
  \mathcal H_{nn}^{\rm LO}
  = G_F\,
  \Big[&
    C_S^{nn}\,(\Lambda^\dagger n)(n^\dagger n)
   +C_T^{nn}\,(\Lambda^\dagger\vec\sigma n)\cdot(n^\dagger\vec\sigma n)
  \Big],
  \label{eq:Hnn}
\end{align}
where
\begin{align}
\label{eq:nnLECs}
&C_{S}^{nn}=(C_{IS}+C_{IV})C_0^0,&C_{T}^{nn}=(C_{IS}+C_{IV})C_0^1.
\end{align}

For $p + n \longrightarrow p + \Lambda$ there are two charged-basis structures,
\begin{align}
\mathcal H_{pn}^{\rm LO}
=G_F\Bigg\{&
(C_{IS}-C_{IV})
\Big[
C_0^0(\Lambda^\dagger n)(p^\dagger p)
 +C_0^1(\Lambda^\dagger\vec\sigma n)\cdot(p^\dagger\vec\sigma p)
\Big]
+2C_{IV}
\Big[
C_0^0(\Lambda^\dagger p)(p^\dagger n)
+C_0^1(\Lambda^\dagger\vec\sigma p)\cdot(p^\dagger\vec\sigma n)
\Big]
\Bigg\}.
\label{eq:Hpn_before}
\end{align}
The first term corresponds to the assignment in which the neutron is
converted into the $\Lambda$ while the proton is the spectator.  The second term
is the charge-exchange assignment present in the isovector piece of the
$\Delta I=1/2$ operator. However, the latter piece can be simplified and rewritten in terms of pure proton spectator terms by using Fierz identities,
\begin{align}
\label{eq:Fierz} &\delta_{\alpha\beta}\delta_{\gamma\delta}=\frac12\delta_{\alpha\delta}\delta_{\gamma\beta}+\frac12\sigma^i_{\alpha\delta}\sigma^i_{\gamma\beta},&
\sigma^i_{\alpha\beta}\sigma^i_{\gamma\delta}=\frac32\delta_{\alpha\delta}\delta_{\gamma\beta}-\frac12\sigma^i_{\alpha\delta}\sigma^i_{\gamma\beta}.
\end{align}
Thus, we obtain
\begin{align}
\label{eq:Hpn}
\mathcal H_{pn}^{\rm LO}
  =G_F\left[
     C_S^{pn}(\Lambda^\dagger n)(p^\dagger p)
    +C_T^{pn}(\Lambda^\dagger\vec\sigma n)\cdot(p^\dagger\vec\sigma p)
  \right],
  \end{align}
where
\begin{align}
\label{eq:pnLECs}
&C_S^{pn}=(C_{IS}-2C_{IV})C_0^0-3C_{IV}C_0^1,
&C_T^{pn}
  =C_{IS}C_0^1-C_{IV}C_0^0. 
\end{align}
Using now the LO fits reported in~\cite{Parreno:2003ny}, $C_0^0=-1.54$, $C_0^1=-0.87$, $C_{IS}=5.01$ and $C_{IV}=1.47$, one derives the LECs in the charge basis used in Eqs.~\eqref{eq:Hnn} and \eqref{eq:Hpn}, arriving at the values shown in Eq.~\eqref{eq:LECsdata} and used in the calculations.  

This formulation is still nonrelativistic: the fields appearing in
Eqs.~\eqref{eq:Hnn} and~\eqref{eq:Hpn} are two-component Pauli baryon fields.
In order to combine the contact contribution coherently with the relativistic OPE and OKE amplitudes, we promote the nonrelativistic operators to covariant four-baryon structures. This covariant embedding is not unique, because the LO nonrelativistic potential fixes only the leading $p/M$ limit of the operator.
In this work we adopt a minimal $VV-AA$ embedding
\begin{align}
(B_c^\dagger B_a)(B_d^\dagger B_b)
&\longrightarrow
\left(\bar u_c\gamma_\mu u_a\right)
\left(\bar u_d\gamma^\mu u_b\right),
\\[1ex]
(B_c^\dagger\vec\sigma B_a)\cdot
(B_d^\dagger\vec\sigma B_b)
&\longrightarrow
-
\left(\bar u_c\gamma_\mu\gamma_5 u_a\right)
\left(\bar u_d\gamma^\mu\gamma_5 u_b\right).
\label{eq:NR_to_cov_dictionary}
\end{align}
The minus sign in the second line ensures that, in the leading
nonrelativistic reduction, the axial--axial current reproduces the $\vec\sigma_1\cdot\vec\sigma_2$ structure of the LO potential. We use the same canonical relativistic normalization of external spinors as in the OPE and OKE
amplitudes.
We now apply this prescription to the channel we took as a concrete example in Eq.~\eqref{eq:app_pn_to_pL}. Using this minimal embedding of Eq.~\eqref{eq:Hpn}, the contact amplitude is
\begin{align}
\mathcal M_{\rm ct}^{pn}
=
-G_F\Big\{
&C_S^{pn}
\left[\bar u_p(p_3)\gamma_\mu u_p(p_1)\right]
\left[\bar u_\Lambda(p_4)\gamma^\mu u_n(p_2)\right]
\nonumber\\
&-
C_T^{pn}
\left[\bar u_p(p_3)\gamma_\mu\gamma_5 u_p(p_1)\right]
\left[\bar u_\Lambda(p_4)\gamma^\mu\gamma_5 u_n(p_2)\right]
\Big\}.
\label{eq:Mct_pn}
\end{align}
Finally, we can put all the pieces together, and perform the spin sums of all the contributions, including the OME terms discussed above
\begin{equation}
\overline{|\mathcal M_{pn}|^2}
=
\frac{1}{4}
\sum_{\rm spins}
\left|
\mathcal M^{\rm ope}_{pn}
+
\mathcal M^{\rm oke}_{pn}
+
\mathcal M_{\rm ct}^{pn}
\right|^2 .
\end{equation}
This expression contains the pure OPE, OKE, and contact pieces, as well as the
OPE--OKE, OPE--contact, and OKE--contact interference terms. For $n + n \longrightarrow n + \Lambda$, the contact amplitude receives two contributions because of the two identical incoming neutrons (see Fig.~\ref{fig:CT})
\begin{equation}
  \mathcal M^{\rm ct}_{nn}
=
\mathcal M^{\rm ct}_{nn,\rm{D}}-\mathcal M^{\rm ct}_{nn,\rm{E}}.
\end{equation}
where the minus sign follows from Wick's theorem.

As discussed above, the covariant embedding of the contact operator is not
unique. A useful way to quantify this ambiguity is to compare the Fierz-rearranged embedding used in Eq.~\eqref{eq:Mct_pn} with an alternative construction in which the original nonrelativistic contact operators in Eq.~\eqref{eq:Hpn_before} are directly promoted to covariant structures following Eq.~\eqref{eq:NR_to_cov_dictionary} without the Fierz rearranging procedure. In the $p + n \longrightarrow p + \Lambda$ channel, this second prescription keeps explicitly the
additional charge-exchange amplitude associated with the second term in the LO
Hamiltonian of Eq.~\eqref{eq:Hpn_before}. Both prescriptions reproduce the same
local nonrelativistic contact operator.
Equivalently, their squared amplitudes agree in the strict nonrelativistic
limit, where the external momenta are small compared with the baryon masses.
Away from this limit, however, the two prescriptions are not equivalent, as a result, the two relativistic amplitudes differ by terms that are subleading in the nonrelativistic expansion but can still be
numerically relevant under PNS conditions.
We have evaluated this effect explicitly in the collision operator for the $p + n \longrightarrow p + \Lambda$ channel using the thermodynamic conditions of Eq.~\eqref{eq:typicalPNS}. The collision operators obtained with the two embeddings differ by about $40\%$. This difference does not change our results and it should be interpreted as an estimate of the systematic uncertainty associated with the relativistic completion of the EFT contact interaction.

It is useful to compare the EFT contact terms with the simple factorized
\(W\)-exchange model used in Ref.~\cite{Alford:2020pld}. In
this model the matrix element of the charged-current interaction is approximated as
\begin{align}
\label{eq:weakHadAmp}
\mathcal M^{W}_{\rm ct}
&=
\frac{G_F \sin2\theta_c}{2\sqrt{2}}\,
\langle\Lambda|
\bar{s}\gamma_\mu (1-\gamma_5)u
|p\rangle\,
\langle p|
\bar u \gamma^\mu (1-\gamma_5)d
|n\rangle
\nonumber\\
&\simeq
\frac{G_F \sin2\theta_c}{2\sqrt{2}}\,
\big[
\bar u_\Lambda
\gamma_\mu
\left(f_1^{\Lambda p}-g_1^{\Lambda p}\gamma_5\right)
u_p
\big]
\big[
\bar u_p
\gamma^\mu
\left(1-g_A\gamma_5\right)
u_n
\big],
\end{align}
where $\sin2\theta_c=2V_{ud}V_{us}$, and
$f_1^{\Lambda p}$ and $g_1^{\Lambda p}$ are the vector and axial-vector
couplings of the $\Lambda \longrightarrow p$ transition. In the non-relativistic limit, the mixed $VA$ and $AV$ terms are suppressed by powers of $p/M$, and only the $VV - AA$ terms survive. This leads to a natural comparison at non-relativistic level with the EFT contact LECs in Eq.~\eqref{eq:pnLECs}. We stress that the comparison holds only at leading order in $p/M$ non-relativistic expansion; nevertheless it provides a qualitative guide for the full result found at the level of collision operators.
Using the same sign convention as in Eq.~\eqref{eq:Mct_pn}, and rearranging the terms in the proton spectator basis through Fierz identities, we find
\begin{equation}
C_{S}^{W,pn}=0.17,
\qquad
C_{T}^{W,pn}=-0.18,
\end{equation}
to be compared with
\begin{equation}
C_S^{pn}=0.54,
\qquad
C_T^{pn}=-2.13
\end{equation}
for the EFT contact interaction used in this work. The numerical impact of this difference is sizeable. Under the same PNS conditions used in Eq.~\eqref{eq:typicalPNS}, replacing the EFT contact
interaction by the $W$-exchange estimate changes the integrated
collision operator for $p + n \longrightarrow p + \Lambda$ by about one order of magnitude. Even though the precise size of the effect is determined by the full phase-space integration, the result is qualitatively consistent with the much smaller coefficient $C_T^{pn}$ obtained from the $W$-exchange model.

\section{Angular integrals and kinematic bounds for semileptonic coalescence rate}
\label{app:main}

In this appendix, we collect the angular kernels $I_{\mathcal A}$, $I_{\mathcal B}$, and $I_{\mathcal K}$ entering the antineutrino absorption rate in Eq.~\eqref{eq:RateSLcoal}. We do not repeat the full derivation here; the interested reader can find it in Appendix B of Ref.~\cite{Guo:2020tgx}, after translating between their notation and ours.

The relevant process is the semileptonic coalescence process
\begin{equation}
  \bar\nu_\ell + B + \ell^- \longrightarrow B',
\end{equation}
for which we denote by $E_X$ and $p_X\equiv|\vec p_X|$ the energy and the three-momentum modulus of the particle $X\in\{\bar\nu_\ell,B,\ell^-,B'\}$.

For the contribution multiplying $\mathcal A$, the kernel is
\begin{align}
  I_{\mathcal A}
  &= -\frac{1}{60}\Bigl[
      3\bigl((p_{\mathcal A,+})^5-(p_{\mathcal A,-})^5\bigr)
      -10(a+b)\bigl((p_{\mathcal A,+})^3-(p_{\mathcal A,-})^3\bigr)
      +60ab\bigl(p_{\mathcal A,+}-p_{\mathcal A,-}\bigr)
      \Bigr]\,,
  \label{eq:app_IA}\\[2mm]
  a &= E_{\bar\nu}E_B+\frac{p_{\bar\nu}^2+p_B^2}{2}\,,\qquad
  b = -E_\ell E_{B'}+\frac{p_\ell^2+p_{B'}^2}{2}\,.
\end{align}
For the contribution multiplying $\mathcal B$, the kernel is
\begin{align}
  I_{\mathcal B}
  &= -\frac{1}{60}\Bigl[
      3\bigl((p_{\mathcal B,+})^5-(p_{\mathcal B,-})^5\bigr)
      -10(c+d)\bigl((p_{\mathcal B,+})^3-(p_{\mathcal B,-})^3\bigr)
      +60cd\bigl(p_{\mathcal B,+}-p_{\mathcal B,-}\bigr)
      \Bigr]\,,
  \label{eq:app_IB}\\[2mm]
  c &= -E_{\bar\nu}E_{B'}+\frac{p_{\bar\nu}^2+p_{B'}^2}{2}\,,\qquad
  d = E_B E_\ell+\frac{p_B^2+p_\ell^2}{2}\,.
\end{align}
For the contribution multiplying $\mathcal K$, the kernel is
\begin{align}
  I_{\mathcal K}
  &= \frac{1}{6}\Bigl[
      \bigl((p_{\mathcal K,+})^3-(p_{\mathcal K,-})^3\bigr)
      +6e\bigl(p_{\mathcal K,+}-p_{\mathcal K,-}\bigr)
      \Bigr]\,,
  \label{eq:app_IC}\\[2mm]
  e &= -E_{\bar\nu}E_\ell-\frac{p_{\bar\nu}^2+p_\ell^2}{2}\,.
\end{align}
The corresponding kinematic bounds are
\begin{align}
  p_{\mathcal A,-} &= \max\left\{|p_{\bar\nu}-p_B|,\ |p_\ell-p_{B'}|\right\},& p_{\mathcal A,+} &= \min\left\{p_{\bar\nu}+p_B,\ p_\ell+p_{B'}\right\},
  \label{eq:app_pA_pm}\\[3pt]
  p_{\mathcal B,-} &= \max\left\{|p_{\bar\nu}-p_{B'}|,\ |p_B-p_\ell|\right\}, &p_{\mathcal B,+} &= \min\left\{p_{\bar\nu}+p_{B'},\ p_B+p_\ell\right\},\\[3pt]
  p_{\mathcal K,-} &= \max\left\{|p_{\bar\nu}-p_\ell|,\ |p_B-p_{B'}|\right\}, &p_{\mathcal K,+} &= \min\left\{p_{\bar\nu}+p_\ell,\ p_B+p_{B'}\right\}.
\end{align}
In all cases the corresponding kernel is non-zero only if $p_{i,-}<p_{i,+}$.

\clearpage
\twocolumngrid
\bibliography{references}
\end{document}